\documentclass[twocolumn]{autart}
\usepackage{mathtools}
\usepackage{amssymb}
\usepackage{balance}
\usepackage{enumitem}
\usepackage{mathdots}
\usepackage{xfrac}
\usepackage{bm}
\usepackage{bbm}
\usepackage{array}
\usepackage{algorithm}
\usepackage{algorithmicx}
\usepackage{algpseudocode}

\usepackage{pgfplots}
\usepgflibrary{shapes}
\usepackage{tikz}

\usetikzlibrary{arrows,calc,patterns}
\tikzstyle{Rect}=[draw=gray,line width=0.001pt,preaction={clip, postaction={pattern=north east lines, pattern color=gray,line width=0.1pt}}]
\tikzset{
	>=stealth',
	help lines/.style={dashed, thick},
	axis/.style={<->},
	important line/.style={thick},
	connection/.style={thick, dotted},
}

\newtheorem{assumption}{Assumption}
\newtheorem{lemma}{Lemma}
\newtheorem{corollary}{Corollary}
\newtheorem{proposition}{Proposition}
\newtheorem{theorem}{Theorem}

\usepackage{etoolbox}

\makeatletter
\def\@xnamedef#1{\expandafter\protected@xdef\csname #1\endcsname}
\def\no@harm{} 
\def\ead@au#1{\protected@edef\@ead@au{#1}}
\patchcmd\runningauthor@fmt{\global\edef}{\protected@xdef}{}{}
\patchcmd\runningauthor@fmt{\global\edef}{\protected@xdef}{}{}
\patchcmd\author@fmt{\edef}{\protected@edef}{}{}
\patchcmd\add@xtok{\xdef}{\protected@xdef}{}{}
\makeatother                                   
\usepackage{color}

\begin{document}
\begin{frontmatter}
\title{Kernel-based identification using Lebesgue-sampled data\thanksref{footnoteinfo}\vspace{-0.1cm}}

\thanks[footnoteinfo]{The material in this paper was partially submitted and accepted for presentation at the 22nd IFAC World Congress (IFAC 2023), July 9-14, 2023, Yokohama, Japan. Corresponding author: R.~A.~González.}

\author[TUE]{Rodrigo A. Gonz\'alez}\ead{r.a.gonzalez@tue.nl},    
\author[TUE]{Koen Tiels}\ead{k.tiels@tue.nl},    
\author[TUE,Delft]{Tom Oomen}\ead{t.a.e.oomen@tue.nl},               
\address[TUE]{Control Systems Technology Section, Department of Mechanical Engineering, Eindhoven University of Technology, Eindhoven, The Netherlands.} 
\address[Delft]{Delft Center for Systems and Control, Delft	University of Technology, Delft, The Netherlands.}  
\vspace{-0.1cm}
\begin{abstract}
Sampling in control applications is increasingly done non-equidistantly in time. This includes applications in motion control, networked control, resource-aware control, and event-based control. Some of these applications, like the ones where displacement is tracked using incremental encoders, are driven by signals that are only measured when their values cross fixed thresholds in the amplitude domain. This paper introduces a non-parametric estimator of the impulse response and transfer function of continuous-time systems based on such amplitude-equidistant sampling strategy, known as Lebesgue sampling. To this end, kernel methods are developed to formulate an algorithm that adequately takes into account the bounded output uncertainty between the event timestamps, which ultimately leads to more accurate models and more efficient output sampling compared to the equidistantly-sampled kernel-based approach. The efficacy of our proposed method is demonstrated through a mass-spring damper example with encoder measurements and extensive Monte Carlo simulation studies on system benchmarks.
\end{abstract}
\end{frontmatter}

\section{Introduction}
In system identification and control design, it is common to assume that the signals are sampled equidistantly in time. However, it is now well known that event-based sampling schemes can lead to improvements in control performance, as well as in resource efficiency \cite{aastrom2003systems}. In particular, one of the most popular event-based sampling methods is Lebesgue sampling. The event associated with this sampling scheme is the crossing of fixed thresholds in the amplitude domain of the continuous-time signal of interest. Such type of sampling can be found in incremental encoders \cite{merry2013optimal}, and also in networked control systems, where the goal is to reduce resource utilization without affecting network throughput \cite{liu2014survey}.

The Lebesgue sampling paradigm provides knowledge on what amplitude band the signals are located in at each instant of time. In this sense, this type of sampling is related to quantization, since a measurement (or lack of) at any instant in time that does not correspond to an event can be viewed as a quantized measurement. There has been extensive work on how to identify systems based on quantized measurements. The maximum likelihood estimator based on the Expectation-Maximization algorithm (EM) has been derived for finite impulse response (FIR) systems in \cite{godoy2011identification}, while \cite{chen2012impulse} develops a regularized FIR estimator for binary measurements. An approximate maximum likelihood approach is studied in \cite{risuleo2019identification}, and \cite{bottegal2017new} proposes a kernel-based method for estimating FIR models. Other approaches have been pursued for the identification of IIR systems \cite{pouliquen2019identification,piga2021learning}, ARX systems \cite{aguero2017based}, and event-based sampling of FIR models with binary observations \cite{diao2018event}.

The problem that is addressed in this paper is the estimation of non-parametric continuous-time models from Lebesgue-sampled output data. To this end, we seek estimators that can 1) provide a \textit{continuous-time} impulse or transfer function estimate from possibly noisy and short data records, and 2) exploit the entirety of the output information contained in the irregular sampling instants and the bounded intersample behavior. Our interest in continuous-time models stems from the fact that they can provide physical interpretability, which is relevant when dealing with applications such as the identification of positioning systems with incremental encoder sensing \cite{strijbosch2022iterative}. Furthermore, direct identification of continuous-time systems can deal with non-uniformly sampled data, which can be the case for event-based sampling schemes, and they can incorporate the full continuous-time input information in the construction of the estimators \cite{gonzalez2021srivc}, which solves the bias problems encountered in discrete-time when the intersample behavior of the input is misspecified \cite{schoukens1994identification}.

Although there has been recent work on non-parametric identification for continuous-time systems using kernel methods that use non-equidistantly sampled data \cite{pillonetto2010new,scandella2022kernel}, these works do not incorporate the intersample behavior information provided by a Lebesgue sampling framework, i.e., the lower and upper bounds on the unsampled output in between the time-stamps are not exploited. In \cite{kawaguchi2016system,pouliquen2016continuous}, continuous-time systems with Lebesgue-sampled and binary outputs are considered, although such results are only valid for parametric models with fixed model structures. On the other hand, the approaches in \cite{chen2012impulse,risuleo2019identification,bottegal2017new} for identification with quantized data might be used for obtaining a non-parametric discrete-time representation that can later be converted into continuous-time. However, this conversion is in many cases ill-defined or ill-conditioned, which drives the need for directly estimating a continuous-time system from the input-output data \cite{garnier2014advantages}.

In summary, the main contributions of this paper are:
\begin{enumerate}[label=(C\arabic*)]
	\item We introduce a loss function (in terms of the continuous-time impulse response to be estimated) that incorporates the intersample information we obtain through Lebesgue sampling. This loss function, after regularization, has an optimum that can be characterized by the generalized representer theorem \cite{wahba1990spline,scholkopf2001generalized}, and is related to a maximum \textit{a posteriori} (MAP) optimization problem for Lebesgue-sampled data.
	
	\item Once the kernel-regularized estimator is written as a finite linear combination of representers, we propose an iterative procedure that delivers the associated weights based on the MAP Expectation-Maximization (MAP-EM) method. We also contrast this procedure with a midpoint approach for identification with quantized data \cite{risuleo2019identification}. 
	
	\item The hyperparameters that describe the kernel and noise variance are computed from an Empirical Bayes (EB) approach. We make the high-dimensional integral optimization problem tractable~by
	\begin{enumerate}[label=(C3.\arabic*)]
		\item Providing closed-form expressions for the kernel matrix in terms of the input samples and the kernel hyperparameters, which is made explicit for the stable-spline kernels, and
		\item Proposing an EM algorithm that iteratively computes the optimal hyperparameter vector. Such algorithm is presented in a matrix-inversion-free form by leveraging Cholesky and QR factorizations. While the noise variance estimate has a closed-form expression for its iterations, the other two hyperparameters are computed via a simple non-convex optimization step.
	\end{enumerate}
	\item We obtain a closed-form expression for the estimated continuous-time transfer function in terms of the representer weight vector, the input samples, and an integrated version of the kernel in the frequency domain. 
	
	\item The proposed method is tested via Monte Carlo simulations.
\end{enumerate}

The remainder of the paper is organized as follows. In Section \ref{sec:problemformulation}, the problem of interest is stated, and practical aspects of Lebesgue-sampled system identification are covered. Section \ref{sec:kernel} introduces the ideas and notation behind non-parametric continuous-time system identification using kernel methods. Section \ref{sec:nonparametricLebesgue} contains the main contribution of this paper, namely, the derivation of a kernel-based estimator for continuous-time, linear and time-invariant (LTI), Lebesgue-sampled systems. Numerical studies are presented in Section \ref{sec:simulations}, while Section \ref{sec:conclusions} provides concluding remarks.

Preliminary results related to the current manuscript are presented in \cite{gonzalez2023impulse}. The present paper substantially extends these results by 1) providing a MAP interpretation to the novel cost function being minimized for identification, 2) introducing an initialization for the MAP-EM approach, 3) proposing more computationally efficient optimization problems for the hyperparameters and 4) deriving the estimated transfer function description in closed form. Additional simulation setups are tested and presented in this paper, and all proofs can be found in the Appendix.

\section{Setup and problem formulation}
\label{sec:problemformulation}

\subsection{System and setup}
Consider the following LTI, asymptotically stable, strictly causal, continuous-time system
\begin{equation}
	\label{system}
	x(t) = \int_{0}^\infty g(\tau) u(t-\tau)\textnormal{d} \tau,
\end{equation}
where $u$ is the input, which is assumed to be a causal function in $t$ (i.e., $u(t)=0$ for $t<0$) that is deterministic and exogenous, and $g$ is the impulse response of the LTI system. The transfer function of the LTI system, defined as the Laplace transform of the impulse response $g$, is denoted as $G(s)$, where $s$ denotes the Laplace complex variable. The frequency response function associated with $g$ is given by evaluating $G(s)$ at $s = \mathrm{i}\omega$.

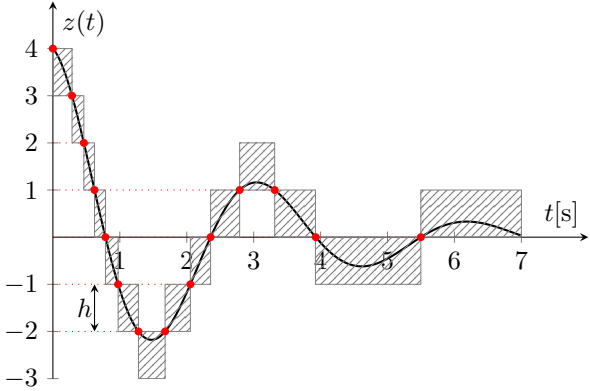
\begin{figure}
	\begin{center}
		\begin{tikzpicture}
			[
			declare function={
				func1(\x)= 	4*exp(-\x/2.5)*cos(deg(2*\x));}
			]
			\begin{axis}[
				width=8.7cm, height=6.6cm, axis x line=middle, axis y line=middle,
				ymin=-3, ymax=5, ytick={-3,...,4}, ylabel=$z(t)$,
				xmin=0, xmax=8, xtick={0,...,7}, xlabel=$t${[s]},
				domain=-1:8,samples=301, 
				]
				\addplot [black,thick,domain=0:7] {func1(x)};
				\addplot [dotted,red] coordinates {(0,3) (0.2859,3)};
				\addplot [dotted,red] coordinates {(0,2) (0.4626,2)};
				\addplot [dotted,red] coordinates {(0,1) (3.314,1)};
				\addplot [dotted,red] coordinates {(0,0) (5.4977,0)};
				\addplot [dotted,red] coordinates {(0,-1) (2.0538,-1)};
				\addplot [dotted,red] coordinates {(0,-2) (1.6765,-2)};
				\draw[Rect] (0,600) rectangle (28.59,700);
				\draw[Rect] (28.59,500) rectangle (46.26,600);
				\draw[Rect] (46.26,400) rectangle (62.22,500);
				\draw[Rect] (62.22,300) rectangle (78.54,400);
				\draw[Rect] (78.54,200) rectangle (97.45,300);
				\draw[Rect] (97.45,100) rectangle (127.83,200);
				\draw[Rect] (127.83,0) rectangle (167.65,100);
				\draw[Rect] (167.65,100) rectangle (205.38,200);
				\draw[Rect] (205.38,200) rectangle (235.62,300);
				\draw[Rect] (235.62,300) rectangle (279.04,400);
				\draw[Rect] (279.04,400) rectangle (331.40,500);
				\draw[Rect] (331.40,300) rectangle (392.27,400);
				\draw[Rect] (392.27,200) rectangle (549.77,300);
				\draw[Rect] (549.77,300) rectangle (700,400);
				\draw [stealth-stealth](62.22,100) -- (62.22,200);
				\node[text width=3cm] at (205,150) 
				{$h$};
				\addplot [only marks,mark=*,red,mark size=1.3] coordinates {(0,4)(0.2859,3)(0.4626,2)(0.6222,1)(2.7904,1)(3.3140,1)(0.7854,0)(2.3562,0)(0.9745,-1)(3.927,0)(5.4977,0)(2.0538,-1)(1.2783,-2)(1.6765,-2)};
			\end{axis}
		\end{tikzpicture}
	\end{center}
	\caption{Lebesgue sampling of a signal $z(t)$ with threshold amplitude $h = 1$. The red dots indicate the sampling instants and thresholds being crossed, and the dashed gray rectangles show the regions where $z(t)$ is known to be located.}
	\label{fig1}
\end{figure}

The input $u(t)$ is assumed to be perfectly known, i.e., there is no noise in its measurement. The output $x(t)$ is corrupted by additive noise $v(t)$, which results in a continuous-time signal $z(t)=x(t)+v(t)$. Assume that we have access to $N_{\textnormal{L}}$ data points of the Lebesgue-sampled version of $z(t)$, as in Fig. \ref{fig1}. That is, given the threshold amplitude $h>0$ and the continuous-time signal $z(t)$, we have at disposal the sampled sequence $\{y_\textnormal{L}(t_l)\}_{l=1}^{N_\textnormal{L}}$ that satisfies $y_\textnormal{L}(t_l)=z(t_l)$. The sampling times (or time-stamps) $t_l, l=1,2,\dots,N_\textnormal{L}$, are the instants in time at which $z(t)$ crosses a fixed threshold $h m_l$, with $m_l\in \mathbb{Z}$. Formally, we characterize the time-stamps by
\begin{align}
	t_l \hspace{-0.04cm}&= \hspace{-0.04cm}\min \hspace{-0.06cm} \left\{ \tau \hspace{-0.04cm} \hspace{-0.04cm}\in (t_{l-1},\infty)\hspace{-0.04cm}:\hspace{-0.04cm} z(\tau) \hspace{-0.04cm}= \hspace{-0.04cm}mh \textnormal{ for some }m\hspace{-0.04cm}\in \hspace{-0.04cm}\mathbb{Z}\right\}\hspace{-0.04cm}, \notag \\
	m_l &= z(t_l)/h. \notag
\end{align}

Without loss of generality and for simplicity only, we assume that $t_1=0$. The goal is to obtain an estimate of the continuous-time system $g$ using the continuous-time input $\{u(t)\}_{t\in [0,t_{N_\textnormal{L}}]}$ and the Lebesgue-sampled output data $\{y_\textnormal{L}(t_l)\}_{l=1}^{N_\textnormal{L}}$. 

\subsection{Practical framework for Lebesgue-sampled system identification}
Incremental encoders operate on this kind of sampling principle \cite{merry2013optimal}. In practice, a light source emits a beam directed towards a slotted disk or strip, and the output of two light detectors are recorded. These two signals allow the encoder to detect the direction of the rotation. These signals are evaluated at a high sampling rate compared to that of the input sequence, typically generated by a zero-order-hold (ZOH) device \cite{strijbosch2019beyond}. The quantity $h$ represents the uncertainty in the measurements of the incremental encoder, which is inversely proportional to its resolution. In low-resolution incremental encoders, the quantization effect produced by $h$, in conjunction with the non-equidistant nature of the sampling mechanism, can impact the performance and design of iterative learning control \cite{strijbosch2022iterative} or repetitive control \cite{kon2021intermittent}.

With this context in mind, we define $\Delta>0$ as the (equidistant) sampling period of the amplitude detection mechanism. The following assumption is set in place:
\begin{assumption}
	\label{assumption12}
	For every time instant $t=i\Delta$, the lower and upper threshold levels associated with the unsampled output $z(t)$ are known. The lower bound at each time instant $t=i\Delta$ is denoted as $\eta_i$, and it can be deduced unambiguously from $\{y_{\textnormal{L}}(t_l)\}_{l=1}^{N_\textnormal{L}}$. 
\end{assumption}

Thus, a set-valued signal $y(i\Delta)$ can be defined as
\begin{equation}
	\label{qh}
	y(i\Delta) = \mathcal{Q}_h\{z(i\Delta)\}:=[\eta_i,\eta_i+h)
\end{equation}
for $i=0,1,\dots,N$, with $N:=\lfloor t_{N_\textnormal{L}}/\Delta\rfloor +1$. To simplify our notation, we denote $\{z(i\Delta)\}_{i=0}^{N}$ as the vector $\mathbf{z}_{0:N}$, and we define the set describing the output measurements as
\begin{equation}
    \label{mathcaly}
\mathcal{Y}_{1:N}\hspace{-0.08cm}=\hspace{-0.1cm}\big\{\hspace{-0.01cm}[z_1,\dots,z_N]^{\hspace{-0.02cm}\top}\hspace{-0.09cm}\in \hspace{-0.03cm}\mathbb{R}^{N}\hspace{-0.03cm}\colon \hspace{-0.03cm}z_i \hspace{-0.04cm}\in \hspace{-0.03cm}y(i\Delta), i\hspace{-0.03cm}=\hspace{-0.03cm}1,\dots,N\big\}.
\end{equation}
Assumption \ref{assumption12} eradicates possible inconsistencies that could occur if $z(t)$ is tangential to one of the threshold levels. Note that we do not assume that each time-stamp $t_l$ is a multiple of $\Delta$. Although such assumption is commonly used in intermittent sampling setups \cite{kon2021intermittent}, and is well justified if the sampling period $\Delta$ is sufficiently small, we do not require it for the proposed method.

\begin{assumption}
	\label{assumption13}
	The sampled disturbance $v(i\Delta)$ affecting the output $z(i\Delta)$ is an additive discrete-time independent and identically distributed (i.i.d.) Gaussian noise of zero mean and variance $\sigma^2$.
\end{assumption}
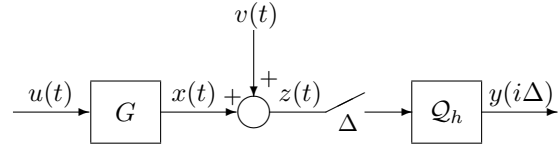
\begin{figure}
	\setlength{\unitlength}{0.1in} 
	\centering 
	\begin{picture}(29,8) 
		\put(0,1.6){\vector(1,0){4}}
		\put(0.8,2.2) {$u(t)$}
		\put(4.1,0){\framebox(3.6,3.2){$G$}}
		\put(7.7,1.6){\vector(1,0){4}}
		\put(8.3,2.2) {$x(t)$}
		\put(11,2.1) {\small{+}}
		\put(11.6,6.3) {$v(t)$}
		\put(12.6,1.6){\circle{1.6}}
		\put(12.9,3) {\small{+}}
		\put(13.4,1.6){\line(1,0){3}}
		\put(13.9,2.2) {$z(t)$}
		\put(12.6,5.9){\vector(0,-1){3.5}}
		\put(16.4,1.6){\line(2,1){2}}
		\put(17,0.6) {\small{$\Delta$}}	
		\put(18.4,1.6){\vector(1,0){2.5}}
		\put(20.9,0){\framebox(3.6,3.2){$\mathcal{Q}_h$}}
		\put(24.5,1.6){\vector(1,0){4}}
		\put(24.9,2.2) {$y(i\Delta)$}
	\end{picture}
	\caption{Block diagram of the Lebesgue sampling scheme. Note that $\mathcal{Q}_h$ delivers a set-valued signal $y$, which is used for identification.}
	\label{fig1:closedloop}
\end{figure}	

The noise variance is not known beforehand, and the user may decide on estimating it from the data or selecting a value according to expert knowledge. For the former approach, it is possible to estimate the noise variance from other tools from identification with quantized measurements \cite{godoy2011identification}, or to include it as an extra hyperparameter to be estimated in the proposed kernel approach. 

\subsection{Problem formulation}

Taking into consideration the Assumptions \ref{assumption12} and \ref{assumption13}, the problem we are interested in is as follows: Assume that the causal continuous-time input $\{u(t)\}_{t\in [0,t_{N_\textnormal{L}}]}$ is perfectly known, i.e., there is no noise in its measurement, and that we have access to $\{y(i\Delta)\}_{i=0}^{N}$, i.e., the upper and lower threshold bounds of $z$. The goal is to estimate the underlying continuous-time impulse response $g$ (or its transfer function $G(s)$) from the input and output data.
\begin{rem}
	\label{remark21}
	In many cases, the input in an identification experiment is generated by a zero-order-hold device of sampling period $\Delta_u$, with $\Delta_u \gg \Delta$. When $\Delta_u$ is a multiple of $\Delta$, we may consider the sampled input signal $\{u(i\Delta)\}_{i=0}^{N}$, instead of a fully continuous-time description for $u$. Clearly both viewpoints describe the same input and are thus equivalent if the intersample behavior of the sampled input is known and correctly incorporated in the construction of the algorithms.
\end{rem}
\begin{rem}
	We will only consider the output data that are produced by the input starting from $t=0$. Since the system to be identified is assumed strictly causal, we discard the first output measurement $y(0)$.
\end{rem}

\section{Kernel-based continuous-time system identification: Preliminaries}
\label{sec:kernel}
This section provides the essential tools behind non-parametric continuous-time system identification using kernel methods, as detailed in, e.g., \cite{pillonetto2014kernel,dinuzzo2015kernels}. In particular, we introduce the notation that is subsequently employed in formulating the proposed non-parametric estimator for Lebesgue-sampled systems.

For unquantized data, estimating the continuous-time impulse response $g$ in a kernel-based framework equates to solving a minimization problem of the form	
\begin{equation}
	\label{regularization}
	\min_{g\in \mathcal{G}} \left(\sum_{i=1}^{N} L\big(z(i\Delta),(g*u)(i\Delta)\big) + \gamma \|g\|_\mathcal{G}^2 \right), 
\end{equation}  
where $\mathcal{G}$ is a Hilbert space of functions, $L(\cdot)$ is a loss function of choice (not necessarily convex \cite{scholkopf2001generalized}), and $\gamma$ is a positive scalar regularization parameter. If the input signal and the space $\mathcal{G}$ are such that all the pointwise evaluated convolutions are bounded linear functionals, then there exist unique representers $\hat{g}_i$ such that $(g*u)(i\Delta)=\langle g,\hat{g}_i \rangle_{\mathcal{G}}$. With this in mind, the representer theorem \cite{scholkopf2001generalized,dinuzzo2012representer} indicates that any optimal solution of \eqref{regularization} can be expressed as a \textit{finite} linear combination of the form 
\begin{equation}
	\label{finiterepresentation}
	\hat{g}(t) = \sum_{i=1}^{N} c_i \hat{g}_i(t),
\end{equation}
where the optimal vector of coefficients $\hat{\mathbf{c}}:=[c_1, c_2,\dots, c_{N}]^\top$ is obtained by
\begin{equation}
	\label{regularizationc}
	\hat{\mathbf{c}}=\underset{\mathbf{c}\in \mathbb{R}^N}{\arg \min}  \left(\sum_{i=1}^{N} L\big(z(i\Delta),\mathbf{K}_i^\top \mathbf{c}\big)+\gamma \mathbf{c}^\top \mathbf{K} \mathbf{c} \right),
\end{equation}
and $\mathbf{K}_i$ denotes the $(i+1)$th column of the kernel matrix $\mathbf{K}$. This matrix is assumed to be non-singular. More explicitly, the representers can be described in terms of the kernel function $k\colon \mathbb{R}_+\times \mathbb{R}_+\to \mathbb{R}$, which fully characterizes the Reproducing Kernel Hilbert Space (RKHS) $\mathcal{G}$. Indeed,
\begin{equation}
	\hat{g}_i(t) = \int_0^\infty u(i\Delta-\tau)k(t,\tau)\textnormal{d}\tau,  \notag
\end{equation}
and the entries of the kernel matrix are given by
\begin{equation}
	\label{kernelmatrix}
	\mathbf{K}_{ij} = \int_0^\infty \int_0^\infty u(i\Delta-\xi) u(j\Delta-\tau)k(\xi,\tau)\textnormal{d}\tau \textnormal{d}\xi.
\end{equation}

One degree of freedom in this framework is the selection of the RKHS space $\mathcal{G}$, which is equivalent to choosing a suitable kernel $k$ with hyperparameters $\bm{\beta}$. There are several kernels for continuous-time impulse response estimation \cite{pillonetto2014kernel}. For example, the stable-spline one of order $q\in\mathbb{N}$ is defined as
\begin{equation}
	k(t,\tau) = s_q(e^{-\beta t},e^{-\beta \tau}), \notag  
\end{equation}
where the hyperparameter $\beta$ is a strictly positive scalar, and $s_q$ is the regular spline kernel of order $q$, given by \cite[Prop. 2.1]{scandella2022kernel}
\begin{equation}
	\label{sq}
	s_q(e^{-\beta t},e^{-\beta \tau}) = \sum_{r=0}^{q-1}\gamma_{q,r} \hspace{-0.03cm}\begin{cases}
		e^{-\beta(2q-r-1)t}e^{-r \beta \tau} & \textnormal{if } t\geq \tau, \\
		e^{-\beta(2q-r-1)\tau}e^{-r \beta t} & \textnormal{if } t<\tau, 	
	\end{cases}
\end{equation}
where
\begin{equation}
	\gamma_{q,r} = \frac{(-1)^{q+r-1}}{r! (2q-r-1)!}. \notag
\end{equation}
In practice, the hyperparameters $\bm{\beta}$, the positive gain $\gamma$ in \eqref{regularization}, and in some cases the noise variance $\sigma^2$, are tuned according to some fitting criteria such as cross validation, the SURE approach \cite{pillonetto2015tuning} or Empirical Bayes \cite{pillonetto2022regularized}. 

\begin{rem}
    The regularization problem in \eqref{regularization} admits a probabilistic interpretation in terms of MAP estimation. Under such perspective, the impulse response $g$ is modeled as a Gaussian process with covariance being described by the kernel $k$. This interpretation is extended in Section \ref{section3A} of this work to the context of Lebesgue sampling. For more details on the probabilistic interpretation for LTI systems, see \cite[Chap. 7]{pillonetto2022regularized}.
\end{rem}

\section{Non-parametric estimation using Lebesgue-sampled data}
\label{sec:nonparametricLebesgue}

In this section, the non-parametric estimator for systems with Lebesgue-sampled data is developed. We divide this section in six parts, which are enumerated next:
\begin{enumerate}
\item The Representer theorem for Lebesgue-sampled systems and its MAP interpretation;
\item A method for initializing the computation of the weights related to each representer;
\item The computation of the optimal weights using the MAP-EM algorithm;
\item The kernel-hyperparameter optimization;
\item A transfer function description for the impulse response estimate; and
\item The full algorithm written in pseudocode.
\end{enumerate}

\subsection{Representer theorem for Lebesgue-sampled systems}
\label{section3A}
The first goal, which constitutes Contribution C1 of this paper, is to derive a loss function $L$ for estimating the impulse response via \eqref{regularization} which incorporates the set knowledge of the output, and to show how it relates with a MAP estimation problem. With that in mind, a Bayesian interpretation of kernel-based methods \cite{kimeldorf1970correspondence,pillonetto2022regularized} involves computing the MAP estimate
\begin{equation}
	\label{mapestimator}
	\hat{g}_{\textnormal{MAP}}(t) = \underset{g}{\arg \max} \big(\ell(g) + \log \textnormal{p}(g)\big),
\end{equation}
where $\textnormal{p}(g)$ is the prior distribution of $g$, which is assumed to be a zero-mean Gaussian process with covariance $\mathbb{E}\{g(t)g(s)\} = k(t,s)/\gamma$, and $\ell(\cdot)$ denotes the log-likelihood function
\begin{equation}
	\ell(g) = \log \textnormal{p}(\mathcal{Y}_{1:N}|g). \notag
\end{equation}
Intuitively, the MAP estimator \eqref{mapestimator} is related to the optimization problem in \eqref{regularization} by letting the\textit{ a priori} probability density of $g$ be proportional to $\exp(-\gamma \|g\|^2_{\mathcal{G}})$, and letting $L$ in \eqref{regularization} be the negative log-likelihood of the measured output data. The main issue that is addressed in this paper is that this argument does not directly hold for $g$ in our case, since the probability density of $g$ is not well defined as it belongs to an infinite-dimensional function space \cite{bogachev1998gaussian}. To this end, a key idea taken here is that it is possible to formalize this intuition by considering the MAP estimator of any \textit{finite} set of samples $(g*u)(t_i)$ that contains the (noiseless) observation set $\{(g*u)(i\Delta)\}_{i=1}^N$. The following lemma uses this insight to provide a formal justification to the choice of $L$ needed for estimating Lebesgue-sampled continuous-time systems.

\begin{lemma}
	\label{lemma31}
	Suppose that Assumptions \ref{assumption12} and \ref{assumption13} hold, and that $g$ is a zero-mean Gaussian process that is independent of $\{v(i\Delta)\}_{i=0}^N$ and has covariance $\mathbb{E}\{g(t)g(s)\} = k(t,s)/\gamma$. Let $\{t_i\}_{i=1}^{N+M}$ be a finite set of real values such that $t_i=i\Delta$ for $i=1,2,\dots,N$, and where $\{t_i\}_{i=N+1}^{N+M}$ are arbitrary. Define the vector of noiseless output values
	\begin{equation}
		\mathbf{x} = [(g*u)(t_1),(g*u)(t_2),\dots,(g*u)(t_{N+M})]^\top. \notag
	\end{equation}
	Furthermore, define $\breve{g}$ as the solution of the optimization problem
	\begin{equation}
		\label{regularization2}
		\min_{g\in \mathcal{G}} \hspace{-0.04cm}\left(\hspace{-0.08cm}-2\hspace{-0.02cm}\sum_{i=1}^{N}\hspace{-0.02cm} \log \hspace{-0.05cm}\left[\hspace{-0.03cm}\int_{\eta_i}^{\eta_i\hspace{-0.03cm}+\hspace{-0.02cm}h}\hspace{-0.1cm} e^{\frac{-1}{2\sigma^2}\big[z_i-(g*u)(i\Delta)\big]^{\hspace{-0.03cm}2}}\hspace{-0.05cm}\textnormal{d}z_i \right] \hspace{-0.1cm}+\hspace{-0.05cm} \gamma \|g\|_\mathcal{G}^2 \right)\hspace{-0.07cm},
	\end{equation}
	where $\|\cdot\|_\mathcal{G}$ is the RKHS norm induced by the kernel $k$. Then, the MAP estimate of $\mathbf{x}$ given $\mathcal{Y}_{1:N}$ is
	\begin{equation}
		\hat{\mathbf{x}} = [(\breve{g} *u)(t_1),(\breve{g}*u)(t_2),\dots,(\breve{g}*u)(t_{N+M})]^\top. \notag
	\end{equation}
\end{lemma}
\begin{pf*}{Proof.}
	For the following analysis, define the first $N$ elements in $\mathbf{x}$ as $\mathbf{x}_1$. The analysis with the first $N$ elements is the relevant and non-standard step, since the MAP estimator of the last $M-N$ elements in $\mathbf{x}$ can be derived with a similar methodology to that in Proposition 5 of \cite{aravkin2014connection}, and is therefore omitted. We first must compute the likelihood function $\log \textnormal{p}(\mathcal{Y}_{1:N}|\mathbf{x}_1)$. To this end, the probability density function of the output prior to sampling $\mathbf{z}_{1:N}=[z(\Delta),\dots,z(N\Delta)]^\top$ (conditioned on $\mathbf{x}_1$) is given by
	\begin{equation}
		\textnormal{p}\left(\mathbf{z}_{1:N}|\mathbf{x}_1\right) = \frac{1}{(2\pi \sigma^2)^{\frac{N}{2}}} \prod_{i=1}^{N} e^{-\frac{1}{2\sigma^2}\big[z(i\Delta)-(g*u)(i\Delta)\big]^2}, \notag
	\end{equation}
	where we have used the fact that the additive noise is Gaussian and i.i.d. by Assumption \ref{assumption13}. Therefore, the probability mass function of $\mathcal{Y}_{1:N}$ is
	\begin{align}
		\hspace{-0.1cm}\textnormal{p}&\hspace{-0.05cm}\left(\mathcal{Y}_{1:N}|\mathbf{x}_1\right) \notag \\
		& \hspace{-0.04cm}= \hspace{-0.03cm} \mathbb{P}\big(z(\Delta)\hspace{-0.05cm}\in \hspace{-0.04cm}[\eta_1,\hspace{-0.02cm}\eta_1\hspace{-0.05cm} +\hspace{-0.05cm} h),\dots,z(N \hspace{-0.03cm} \Delta)\hspace{-0.05cm}\in\hspace{-0.05cm} [\eta_{N},\hspace{-0.02cm} \eta_{N}\hspace{-0.05cm}+\hspace{-0.05cm} h)|\mathbf{x}_1\big) \notag \\
		\label{eq5}
		&\hspace{-0.04cm}= \hspace{-0.03cm} \frac{1}{(2\pi \sigma^2)^{\frac{N}{2}}} \prod_{i=1}^{N}\int_{\eta_i}^{\eta_{i}+h} \hspace{-0.05cm} e^{-\frac{1}{2\sigma^2}\big[z_i-(g*u)(i\Delta)\big]^2}\textnormal{d}z_i.
	\end{align}
	From \eqref{eq5}, the log-likelihood function $\ell(\mathbf{x}_1)$ can be written as
	\begin{equation}
		\ell(\mathbf{x}_1) = \sum_{i=1}^{N} \log \left[\int_{\eta_i}^{\eta_{i}+h} e^{-\frac{1}{2\sigma^2}\big[z_i-(g*u)(i\Delta)\big]^2}\textnormal{d}z_i \right]+C, \notag 
	\end{equation}
	where $C$ is a known constant. On the other hand, $\mathbf{x}_1$ is zero-mean and normally distributed with covariance that has entries given by
	\begin{align}
		\mathbb{E}&\{(g*u)(i\Delta)(g*u)(j\Delta)\} \notag \\
		&= \mathbb{E}\left\{\int_{0}^\infty u(i\Delta-\tau)g(\tau)\textnormal{d}\tau \int_{0}^\infty u(j\Delta-\xi)g(\xi)\textnormal{d}\xi \right\} \notag \\
		&= \int_{0}^\infty \int_{0}^\infty u(i\Delta-\tau)u(j\Delta-\xi)\mathbb{E}\{ g(\tau) g(\xi) \}\textnormal{d}\tau\textnormal{d}\xi \notag \\ 
		\label{satisfies}
		&=\mathbf{K}_{ij}/\gamma. 
	\end{align}
	This leads to the following MAP estimator for $\mathbf{x}_1$:
	\begin{align}
		\hat{\mathbf{x}}_1 &= \underset{\mathbf{x}_1}{\arg \max} \big(\ell(\mathbf{x}_1) + \log\textnormal{p}(\mathbf{x}_1)\big) \notag \\
		&\hspace{-0.49cm}= \hspace{-0.05cm}\underset{\mathbf{x}_1}{\arg \max}\hspace{-0.07cm}\left(\sum_{i=1}^{N}\hspace{-0.04cm} \log \hspace{-0.05cm}\left[\hspace{-0.06cm}\int_{\eta_i}^{\eta_i\hspace{-0.03cm}+\hspace{-0.02cm}h}\hspace{-0.3cm} e^{\hspace{-0.01cm}\frac{-1}{2\sigma^{\hspace{-0.02cm}2}}\hspace{-0.06cm}\big[\hspace{-0.03cm}z_{\hspace{-0.02cm}i}\hspace{-0.04cm}-\hspace{-0.03cm}(\hspace{-0.01cm}g*u\hspace{-0.01cm})\hspace{-0.02cm}(\hspace{-0.02cm}i\Delta\hspace{-0.02cm})\hspace{-0.02cm}\big]^{\hspace{-0.03cm}2}}\hspace{-0.1cm}\textnormal{d}z_{\hspace{-0.02cm}i} \hspace{-0.04cm}\right] \hspace{-0.16cm}-\hspace{-0.1cm} \frac{\gamma\mathbf{x}_{\hspace{-0.02cm}1}^{\hspace{-0.06cm}\top} \hspace{-0.04cm}\mathbf{K}^{\hspace{-0.05cm}-\hspace{-0.03cm}1}\hspace{-0.02cm}\mathbf{x}_1}{2}\hspace{-0.07cm}\right)\hspace{-0.09cm}, \notag
	\end{align}
	Under the representation $(g*u)(i\Delta)=\mathbf{K}_i^\top\mathbf{c}$ for $i=1,2,\dots,N$, we obtain that $\hat{\mathbf{x}}_1 = \mathbf{K}\hat{\mathbf{c}}$, where
	\begin{equation}
		\label{computec}
		\hat{\mathbf{c}} \hspace{-0.03cm}= \hspace{-0.03cm} \underset{\mathbf{c}\in \mathbb{R}^{N}}{\arg \max} \hspace{-0.06cm} \left(\sum_{i=1}^{N}\log \hspace{-0.04cm}\left[\hspace{-0.02cm}\int_{\eta_i}^{\eta_i\hspace{-0.03cm}+\hspace{-0.02cm}h}\hspace{-0.25cm} e^{\frac{-1}{2\sigma^2}\big(\hspace{-0.03cm}z_i\hspace{-0.03cm}-\hspace{-0.03cm}\mathbf{K}_i^\top \mathbf{c}\big)^{\hspace{-0.02cm}2}}\hspace{-0.09cm}\textnormal{d}z_i \right]\hspace{-0.05cm}-\hspace{-0.05cm} \frac{\gamma\mathbf{c}^{\hspace{-0.03cm}\top}\hspace{-0.02cm} \mathbf{Kc}}{2}\hspace{-0.02cm} \right)\hspace{-0.05cm}.
	\end{equation}
	This is precisely the optimal weighting of the representers that describe the solution of \eqref{regularization2} via the representer theorem. This completes the proof. \hfill $\hspace{1cm}$\qed
\end{pf*}
Lemma \ref{lemma31} provides a relation between the impulse response minimization problem in \eqref{regularization} and the MAP estimator in a Lebesgue-sampling framework. More precisely, we have shown that the following choice of loss function for when the output is Lebesgue-sampled
\begin{align}
    L\big(y(i\Delta)&,(g*u)(i\Delta)\big) \notag \\
    &= 2\sum_{i=1}^{N} \log \left[\int_{\eta_i}^{\eta_i+h} e^{\frac{-1}{2\sigma^2}\big[z_i-(g*u)(i\Delta)\big]^{2}}\textnormal{d}z_i \right] \notag
\end{align}
\hspace{-0.15cm}leads to a MAP estimator of the noiseless output of the system prior to Lebesgue sampling. Since the integer $M$ in Lemma \ref{lemma31} is arbitrary, $(\breve{g}*u)(t)$ represents a MAP estimator of the noiseless output for any time instant $t\in[\Delta,\Delta N]$.

The following subsections are focused on how to compute the minimizer of \eqref{regularization2}, and how to choose a specific kernel according to the Lebesgue-sampled data. The optimization problem in \eqref{regularization2} does not have an explicit form as the Riemann sampling counterpart, i.e., the point-valued output case, \cite{pillonetto2022regularized}. However, the representer theorem indicates that any optimal solution of \eqref{regularization2} can anyway be expressed as a \textit{finite} linear combination of the representers $\hat{g}_i$ of the form \eqref{finiterepresentation} with $\hat{\mathbf{c}}$ being given by \eqref{computec}. Next, we cover how to compute $\hat{\mathbf{c}}$, the optimal weighting of the representers $\hat{g}_i$, for a fixed kernel $k$ and hyperparameters $\gamma$ and $\sigma^2$. 

\subsection{Optimal weights with MAP-EM}

In this subsection, we present a MAP-EM algorithm to obtain an iterative procedure that computes \eqref{computec}. The derivation of this iterative procedure, which ensures the computation of a local maximum of the cost in \eqref{computec} under general conditions as a generalization of the standard EM approach \cite{wu1983convergence,mclachlan2007algorithm}, constitutes Contribution C2 of this paper. The approach consists of relating \eqref{computec} to the MAP of a specific FIR model in discrete-time, to later apply the EM algorithm \cite{dempster1977maximum} tailored for MAP estimation. This relation is made evident in the following lemma.
\begin{lemma}
	\label{lemma32}
	Consider the following model
	\begin{subequations}
		\label{modelforc}
		\begin{align}
			\label{modelforc1}
			z(i\Delta) &= \mathbf{K}_i^\top \mathbf{c} + e(i\Delta),  \\
			\label{modelforc2}
			y(i\Delta) &= \mathcal{Q}_h\{z(i\Delta)\}, 
		\end{align}
	\end{subequations}
	where $e(\Delta),\dots, e(N\Delta)$ are i.i.d. Gaussian with variance $\sigma^2$, and $\mathbf{K}_i, i=1,2,\dots,N$, is assumed known. Assume that $\mathbf{c}$ in \eqref{modelforc1} has a Gaussian prior distribution, with zero mean and covariance $(\gamma\mathbf{K})^{-1}$. Then, the MAP estimator for $\mathbf{c}$ is given by $\hat{\mathbf{c}}$ in \eqref{computec}.
\end{lemma}
\begin{pf*}{Proof.}
	See Appendix \ref{prooflemma32}. \hfill $\hspace{1cm}$\qed
\end{pf*}

By Lemma \ref{lemma32} we can view the computation of the weights $\hat{\mathbf{c}}$ in a MAP-EM framework if we set the unquantized data $\mathbf{z}_{1:N}$ as our hidden variable. In other words, we can optimize the\textit{ a posteriori} density for $\mathbf{c}$, which is exactly the objective function in \eqref{computec}, by iteratively 1) computing the conditional expectation of the log complete-data posterior density given the set measurements $\mathcal{Y}_{1:N}$ and the current estimate of $\hat{\mathbf{c}}$ (i.e., the E-step), and later 2) performing a maximization step (M-step). These two steps are outlined in Algorithm \ref{algorithm1}.  Note that this method departs from the standard EM method in the objective function of the maximization step, which here includes the log prior density. The E-step is computed using a result from quantized FIR maximum likelihood estimation, while the M-step including the log prior density is presented in Theorem~\ref{theorem31}.

\begin{algorithm}
	\caption{MAP-EM algorithm for the computation of $\hat{\mathbf{c}}$ in \eqref{computec}}
	\begin{algorithmic}[1]
		\State Select an initial estimate $\hat{\mathbf{c}}^{(1)}$, a maximum number of iterations $M_{\textnormal{iter}}$, and a tolerance factor $\epsilon$
          \State $j\gets 1$, $\textnormal{flag}\gets 1$
		\While{$j\leq M_{\textnormal{iter}}$ and $\textnormal{flag}=1$}
		\State \textbf{E-step}: Compute the expectation
		\begin{equation}
			\label{qfunction}
			Q(\mathbf{c},\hspace{-0.02cm}\hat{\mathbf{c}}^{(j)}\hspace{-0.01cm}) \hspace{-0.04cm}= \hspace{-0.04cm}\mathbb{E}\hspace{-0.03cm}\left\{ \log \textnormal{p}(\mathbf{z}_{1:N}\hspace{-0.02cm},\hspace{-0.02cm}\mathcal{Y}_{1:N}|\mathbf{c}\hspace{-0.02cm})|\mathcal{Y}_{1:N}\hspace{-0.02cm},\hspace{-0.02cm}\hat{\mathbf{c}}^{(j)}\hspace{-0.04cm}\right\}\hspace{-0.03cm}.\hspace{-0.1cm}
		\end{equation} 
		\State \textbf{M-step}: Solve the optimization problem
		\begin{equation}
			\label{mstep}
			\hat{\mathbf{c}}^{(j+1)} = \underset{\mathbf{c}\in \mathbb{R}^N}{\arg \max} \left( Q(\mathbf{c},\hat{\mathbf{c}}^{(j)}) -\frac{\gamma\mathbf{c}^{\top}\mathbf{Kc}}{2} \right).
		\end{equation}
		\If{$\dfrac{\|\hat{\mathbf{c}}^{(j+1)}-\hat{\mathbf{c}}^{(j)}\|_2}{\|\hat{\mathbf{c}}^{(j)}\|_2}<\epsilon$} 		\State $\textnormal{flag} \gets 0$
		\EndIf
		\State $j \gets j+1$
		\EndWhile
	\end{algorithmic}
	\label{algorithm1}
\end{algorithm}

\begin{lemma}{\cite[Lemma 5]{godoy2011identification}}.
	\label{lemma33}
	Consider the discrete-time model \eqref{modelforc}. The $Q$ function in \eqref{qfunction} satisfies
	\begin{align}
		Q&(\mathbf{c},\hat{\mathbf{c}}^{(j)}) \hspace{-0.03cm}= \notag \\
		&\frac{-1}{2\sigma^2} \sum_{i=1}^{N}\int_{\eta_i}^{\eta_i+h} (z_i-\mathbf{K}_i^\top \mathbf{c})^{2} \textnormal{p}(z_i|y(i\Delta),\hat{\mathbf{c}}^{(j)})\textnormal{d}z_i+C, \notag
	\end{align}
where $C$ is a constant.
\end{lemma}
\begin{pf*}{Proof.}
	See \cite{godoy2011identification}. \hfill $\hspace{1cm}$\qed
\end{pf*}
\begin{theorem}
	\label{theorem31}
	The M-step in \eqref{mstep} is equivalent to
	\begin{equation}
		\label{ck1}
		\hat{\mathbf{c}}^{(j+1)} = (\mathbf{K}+ \tilde{\gamma} \mathbf{I})^{-1} \tilde{\mathbf{z}}^{(j)},
	\end{equation}
	where $\tilde{\gamma} = \gamma\sigma^2$, and with the $i$th entry of $\tilde{\mathbf{z}}^{(j)}$ being given~by
	\begin{equation}
		\label{tildezi}
		\tilde{z}_i^{(j)} \hspace{-0.1cm} = \hspace{-0.03cm}\mathbf{K}_i^{\hspace{-0.04cm}\top} \hspace{-0.04cm}\hat{\mathbf{c}}^{\hspace{-0.02cm}(j)} \hspace{-0.04cm}+ \frac{\hspace{-0.16cm}\sqrt{\hspace{-0.07cm}\frac{2}{\pi}}\sigma \hspace{-0.08cm}\left(\hspace{-0.07cm}\exp\hspace{-0.06cm}\big\{\hspace{-0.12cm}-\hspace{-0.09cm}(\hspace{-0.02cm}b_i^{\hspace{-0.03cm}(j)}\hspace{-0.03cm})^2 \hspace{-0.04cm}\big\} \hspace{-0.07cm}-\hspace{-0.04cm} \exp\hspace{-0.06cm}\big\{\hspace{-0.12cm}-\hspace{-0.09cm}(\hspace{-0.02cm}b_i^{\hspace{-0.03cm}(j)}\hspace{-0.15cm}+\hspace{-0.1cm}\frac{h}{\sqrt{\hspace{-0.02cm}2} \sigma}\hspace{-0.03cm})^2 \hspace{-0.04cm}\big\} \hspace{-0.07cm} \right)}{\textnormal{erf}\big[b_i^{(j)}\hspace{-0.07cm}+\hspace{-0.07cm}\frac{h}{\sqrt{2} \sigma}\big]-\textnormal{erf}\big[b_i^{(j)}\big]}\hspace{-0.02cm},
	\end{equation}
	where $b_i^{(j)} := (\eta_i-\mathbf{K}_i^\top \hat{\mathbf{c}}^{(j)})/(\sqrt{2} \sigma)$, and the error function $\textnormal{erf}[x]$ is defined by
	\begin{equation}
		\textnormal{erf}[x] = \frac{2}{\sqrt{\pi}} \int_{0}^{x} e^{-t^2}\textnormal{d}t. \notag
	\end{equation}
\end{theorem}
\begin{pf*}{Proof.}
	See Appendix \ref{prooftheorem31}. \hfill $\hspace{1cm}$\qed
\end{pf*}

Theorem \ref{theorem31} reveals that the optimal weights $\hat{\mathbf{c}}$ can be computed from successive regularized least squares expressions. These have the same form as the standard solution for the optimal weights for unquantized data \cite[Theorem 7.3]{pillonetto2022regularized}, but with an iteration-varying output vector $\tilde{\mathbf{z}}^{(j)}$. Interestingly, $\tilde{z}_i^{(j)}$ can be interpreted as the conditional mean of $z(i\Delta)$ given the available quantized data and the current weight vector $\hat{\mathbf{c}}^{(j)}$; see Eq. \eqref{intermediate0} of Appendix \ref{prooftheorem31} for this interpretation.

\begin{rem}
The iterations provided by the M-step in Theorem \ref{theorem31} require an initial estimate $\hat{\mathbf{c}}^{(1)}$. To this end, by noting that $\tilde{z}_i^{(j)} \in [\eta_i,\eta_i+h)$ for all $i=1,2,\dots, N$, we may follow a best worst-case approach and set 
\begin{equation}
	\label{approximatec}
	\hat{\mathbf{c}}^{(1)} = (\mathbf{K}+\tilde{\gamma}\mathbf{I})^{-1} \tilde{\mathbf{z}}^{(0)},
\end{equation}
with the $i$th entry of $\tilde{\mathbf{z}}^{(0)}$ being the midpoints of each quantization level, i.e., $\tilde{z}_i^{(0)}=\eta_i+h/2$. This initialization coincides with the approach suggested in \cite{risuleo2019identification} for constructing an approximate maximum likelihood estimator under quantized data. 
\end{rem}

\subsection{Kernel hyper-parameter optimization}
\label{subD}

Here we consider the marginal likelihood method for computing an appropriate hyperparameter vector, also known as the Empirical Bayes approach. This approach, which has been proven useful in other contributions on kernel system identification \cite{pillonetto2010new,pillonetto2014kernel,bottegal2017new,scandella2022kernel}, proposes to estimate the hyperparameter vector $\bm{\rho}=[\bm{\beta}^\top, \gamma, \sigma^2]^\top$ by solving the maximum likelihood problem
\begin{equation}
	\label{eb}
	\hat{\bm{\rho}}_{\textnormal{EB}} = \underset{\bm{\rho}\in \bm{\Gamma}}{\arg \max} \hspace{0.1cm}\textnormal{p}(\mathcal{Y}_{1:N}|\bm{\rho}),  
\end{equation}
where $\bm{\Gamma}$ denotes the admissible space of hyperparameters, which must consider $\gamma, \sigma^2>0$. To describe such optimization problem more explicitly, we first compute the probability density function of the output prior to Lebesgue sampling. This expression can be obtained directly by exploiting the fact that the additive noise is Gaussian and independent of $g$ (which is also assumed Gaussian, and satisfies \eqref{satisfies}), thus leading to
\begin{equation}
	\label{zrho}
	\mathbf{z}_{1:N}|\bm{\rho} \sim \mathcal{N}(\mathbf{0},\mathbf{K}_{\bm{\beta}}/\gamma+\sigma^2 \mathbf{I}),
\end{equation}
where we have made explicit the dependence of the kernel matrix $\mathbf{K}$ on the kernel hyperparameter vector $\bm{\beta}$. Therefore, the Empirical Bayes estimator for $\bm{\rho}$ is given by
\begin{align}
	\hat{\bm{\rho}}_{\textnormal{EB}} &= \underset{\bm{\rho} \in \bm{\Gamma}}{\arg \max} \frac{1}{\sqrt{\det(2\pi[\mathbf{K}_{\bm{\beta}}/\gamma+\sigma^2 \mathbf{I}])}} \notag \\
	\label{kernelopt}
	&\times \int_{\mathbf{z}\in \mathcal{Y}_{1:N}} \hspace{-0.32cm}\exp\left\{-\frac{1}{2}\mathbf{z}^\top (\mathbf{K}_{\bm{\beta}}/\gamma + \sigma^2 \mathbf{I})^{-1}\mathbf{z} \right\} \textnormal{d}\mathbf{z},
\end{align}
where $\mathcal{Y}_{1:N}$ is defined in \eqref{mathcaly}. This non-convex optimization problem involves an $N$-dimensional integral, which is hard to compute in general (see, e.g., \cite{chen2012impulse,bottegal2017new}). The intractability is here solved by optimizing \eqref{kernelopt} with EM along similar lines as in the previous subsection. For brevity, we derive the EM iterations jointly (both E and M steps) in Theorem \ref{theorem32}.
\begin{theorem}
	\label{theorem32}
	The following iterative procedure is guaranteed to converge with probability 1 to a (local or global) maximum for the cost in \eqref{kernelopt}:
	\begin{equation}
		\label{emiterationsthm32}
		\hat{\bm{\rho}}^{(j+1)} = \underset{\bm{\rho} \in \bm{\Gamma}}{\arg \min} \left(\log \det(\mathbf{S}_{\bm{\rho}})+\textnormal{tr}\{\mathbf{S}_{\bm{\rho}}^{-1} \bar{\mathbf{Q}}^{(j)}\}\right),
	\end{equation}
	where $\mathbf{S}_{\bm{\rho}}:= \mathbf{K}_{\bm{\beta}}/\gamma + \sigma^2 \mathbf{I}$, and $\bar{\mathbf{Q}}^{(j)}$ is the second moment of $\mathbf{z}_{1:N}$ given the data and the $j$th iteration of $\hat{\bm{\rho}}$, i.e.,
	\begin{equation}
		\label{secondmoment}
		\bar{\mathbf{Q}}^{(j)} = \mathbb{E}\{\mathbf{z}_{1:N}\mathbf{z}_{1:N}^\top |\mathcal{Y}_{1:N},\hat{\bm{\rho}}^{(j)}\}.  
	\end{equation}
\end{theorem}
\begin{pf*}{Proof.}
	See Appendix \ref{prooftheorem32}. \hfill $\hspace{1cm}$\qed
\end{pf*}

\begin{rem}
	The $\bar{\mathbf{Q}}^{(j)}$ matrix in \eqref{secondmoment} cannot be computed in closed-form in general. In this paper, we extract samples of a multivariate truncated Gaussian distribution using the minimax tilting algorithm in \cite{botev2017normal} and we approximate the expectation in \eqref{secondmoment} via Monte Carlo integration.
\end{rem}

The iterations in \eqref{emiterationsthm32} to solve \eqref{kernelopt} can possibly be ill-conditioned and computationally costly to compute. In particular, the kernel matrix $\mathbf{K}$, with elements described in \eqref{kernelmatrix}, is known to be difficult to compute for continuous-time system identification due to the presence of integrals instead of sums in the discrete-time case \cite{dinuzzo2015kernels,scandella2022kernel}. Here we provide the necessary details to explicitly write the elements of this matrix for any kernel $k$ in terms of samples of an input with zero-order hold intersample behavior (recall Remark \ref{remark21}), which is later used in Theorem \ref{theorem33} for constructing more computationally efficient iterations for solving \eqref{kernelopt}. The following lemma and its corollary (Corollary \ref{coroO}) constitute Contribution C3.1 of the paper.

\begin{lemma}
	\label{lemmaO}
	Consider the kernel matrix $\mathbf{K}$ with entries described in \eqref{kernelmatrix}. If $u(t)$ is constant between the time instants $t=0,\Delta, 2\Delta, \dots, N\Delta$, then $\mathbf{K}$ admits the decomposition
	\begin{equation}
		\label{Kdecomposition}
		\mathbf{K}_{\bm{\beta}} = \bm{\Phi} \mathcal{O}_{\bm{\beta}} \bm{\Phi}^\top, 
	\end{equation}
	where $\bm{\Phi}$ is given by
	\begin{equation}
		\label{U}
		\bm{\Phi}=\begin{bmatrix}
			u(0) & & & 0 \\
			u(\Delta) & u(0) & & \\
			\vdots & & \ddots & \\
			u([N\hspace{-0.04cm}-\hspace{-0.04cm}1]\Delta) & u([N\hspace{-0.04cm}-\hspace{-0.04cm}2]\Delta) & \cdots & u(0) 
		\end{bmatrix},
	\end{equation}
	and the matrix $\mathcal{O}_{\bm{\beta}} \in \mathbb{R}^{N\times N}$ has entries
	\begin{equation}
		\label{Oij}
		\mathcal{O}_{\bm{\beta},ij} = \int_{\Delta [i-1]}^{\Delta i} \int_{\Delta [j-1]}^{\Delta j} k(\xi,\tau) \textnormal{d}\tau \textnormal{d}\xi.
	\end{equation}
\end{lemma}
\begin{pf*}{Proof.}
	See Appendix \ref{appendixlemmaO}. \hfill $\hspace{1cm}$\qed
\end{pf*}
\begin{corollary}
	\label{coroO}
	Consider the kernel matrix $\mathbf{K}$ with entries described in \eqref{kernelmatrix}, with $k$ being the stable-spline kernel of order $q$ in \eqref{sq}. If $u(t)$ is constant between the time instants $t=0,\Delta, 2\Delta, \dots, N\Delta$, then $\mathbf{K}$ admits the decomposition $\mathbf{K}_{\beta} = \bm{\Phi} \mathcal{O}_\beta \bm{\Phi}^\top$,
	where $\bm{\Phi}$ is given by \eqref{U} and the matrix $ \mathcal{O}_\beta \in \mathbb{R}^{N\times N}$ has entries
	\begin{equation}
		\mathcal{O}_{\beta, ij} = \sum_{r=0}^{q-1} \frac{\gamma_{q,r} e^{-\beta \Delta (2q-1)\max\{i,j\}}}{\beta^2 r(2q-r-1)}\begin{cases}
			a(\beta) & \textnormal{if } i = j,  \\
			b_{i-j}(\beta) & \textnormal{if }i \neq j,
		\end{cases} \notag
	\end{equation}
	where
	\begin{align}
		a(\beta) &= \frac{2[(2q-r-1)+re^{\beta \Delta (2q\hspace{-0.02cm}-\hspace{-0.02cm}1)}-(2q-1)e^{\beta \Delta r}]}{(2q-1)}, \notag \\
		b_{i-j}(\beta)  &= e^{-\beta \Delta r(1-|i-j|)}(e^{\beta r \Delta}-1)(e^{\beta \Delta(2q-1)}- e^{\beta\Delta r}). \notag
	\end{align}
\end{corollary}
\begin{pf*}{Proof.}
	Direct from replacing $k(\xi,\tau)$ in \eqref{Oij} for \eqref{sq} and solving the integrals. \hfill $\hspace{1cm}$\qed
\end{pf*}

\begin{rem}
	The continuous-time setting provides substantial freedom compared to discrete-time approaches for incorporating the intersample behavior of the input signal. Although Lemma \ref{lemmaO} and Corollary \ref{coroO} are exact only for zero-order hold inputs, these results can be extended in exact form (at the expense of more computations but avoiding numerical integration techniques), to any input with a specified intersample behavior (e.g., first-order hold, or B-splines used in a generalized hold framework \cite{arriagada2008relationship}). Throughout this paper, only ZOH is considered; extensions to other interpolation schemes are conceptually straightforward.
\end{rem}
The description for $\mathbf{K}$ in Lemma \ref{lemmaO} is now used to rewrite the iterations in \eqref{emiterationsthm32} by considering an adequate QR factorization of the data at hand. For the following, we consider the change of variable $\tilde{\gamma} = \gamma\sigma^2$ and compute the Cholesky factorizations $\mathcal{O}_{\bm{\beta}}/\tilde{\gamma} = \mathbf{L}_{\bm{\rho}} \mathbf{L}_{\bm{\rho}}^\top$ and $\bar{\mathbf{Q}}^{(j)}=\mathbf{C}^{(j)}{\mathbf{C}^{(j)}{}}^{\top}$, where $\mathbf{L}_{\bm{\rho}}$ and $\mathbf{C}^{(j)}$ are upper triangular matrices with non-negative diagonal entries. We introduce the QR factorization
\begin{equation}
	\label{qr}
	\begin{bmatrix}
		\bm{\Phi} \mathbf{L}_{\bm{\rho}} & \mathbf{C}^{(j)} \\
		\mathbf{I} & \mathbf{0}
	\end{bmatrix} = \mathbf{Q}_{\bm{\rho}}  \begin{bmatrix}
		\mathbf{R}_{1,\bm{\rho}} & \mathbf{R}_{2,\bm{\rho}} \\
		\mathbf{0} & \mathbf{R}_{3,\bm{\rho}}
	\end{bmatrix},
\end{equation}
where $\mathbf{Q}_{\bm{\rho}}$ is an orthogonal matrix (not to be confused with $\bar{\mathbf{Q}}^{(j)}$ in \eqref{secondmoment}), and $\mathbf{R}_{1,\bm{\rho}}$, $\mathbf{R}_{3,\bm{\rho}}$ are upper triangular matrices of dimension $N\times N$. Without loss of generality, we assume that they have positive diagonal entries. Note that the following identities are satisfied:
\begin{subequations}
	\label{Rs}
	\begin{align}
		\label{R1R1}
		\mathbf{R}_{1,\bm{\rho}}^\top \mathbf{R}_{1,\bm{\rho}} &=  \mathbf{L}_{\bm{\rho}}^\top \bm{\Phi}^\top \bm{\Phi} \mathbf{L}_{\bm{\rho}} + \mathbf{I},  \\
		\label{R1R2}
		\mathbf{R}_{1,\bm{\rho}}^\top \mathbf{R}_{2,\bm{\rho}} &=  \mathbf{L}_{\bm{\rho}}^\top \bm{\Phi}^\top \mathbf{C}^{(j)}, \\
		\mathbf{R}_{2,\bm{\rho}}^\top \mathbf{R}_{2,\bm{\rho}} + \mathbf{R}_{3,\bm{\rho}}^\top \mathbf{R}_{3,\bm{\rho}} &= {\mathbf{C}^{(j)}{}}^\top \mathbf{C}^{(j)}.  
	\end{align}
\end{subequations}
Theorem \ref{theorem33} provides a straightforward implementation for computing the EM iterations of Theorem \ref{theorem32}, which constitutes Contribution C3.2 of this paper.

\begin{theorem}
	\label{theorem33}
	The iterative procedure in \eqref{emiterationsthm32} for computing $\hat{\bm{\rho}}_{\textnormal{EB}}$ in \eqref{kernelopt} is equivalent to
	\begin{align}
		&\begin{bmatrix}
			\hat{\tilde{\gamma}}^{(j+1)}  \\
			\hat{\bm{\beta}}^{(j+1)}
		\end{bmatrix} \notag \\
		\label{lambdabetaopt}
		&\hspace{-0.3cm}=  \underset{\tilde{\gamma},\bm{\beta}}{\arg \min} \hspace{-0.04cm}\left(\hspace{-0.03cm}N\hspace{-0.02cm}\log \hspace{-0.02cm}\left(\hspace{-0.02cm}\|\mathbf{C}^{(j)}\hspace{-0.03cm}\|_{\textnormal{F}}^2 \hspace{-0.03cm}-\hspace{-0.05cm}\|\mathbf{R}_{2,\bm{\rho}}\|_{\textnormal{F}}^2\right)\hspace{-0.05cm} + \hspace{-0.02cm} 2\hspace{-0.02cm}\log \hspace{-0.02cm}\det (\hspace{-0.01cm}\mathbf{R}_{1\hspace{-0.02cm},\bm{\rho}}\hspace{-0.01cm})\hspace{-0.02cm}\right)\hspace{-0.05cm}, \\
		\label{sigmaopt}
		&{\hat{\sigma}^2{}}^{(j+1)} = \frac{1}{N} \left(\|\mathbf{C}^{(j)}\|_{\textnormal{F}}^2-\|\mathbf{R}_{2,\hat{\bm{\rho}}^{(j+1)}}\|_{\textnormal{F}}^2\right),
	\end{align}
	where $\|\cdot\|_\textnormal{F}$ is the Frobenius norm, $\mathbf{R}_{1,\bm{\rho}}$ and $\mathbf{R}_{2,\bm{\rho}}$ are computed from \eqref{qr}, and $\mathbf{C}^{(j)}$ is the Cholesky factor of $\bar{\mathbf{Q}}^{(j)}$ in \eqref{secondmoment}.
\end{theorem}
\begin{pf*}{Proof.}
	See Appendix \ref{prooftheorem33}. \hfill $\hspace{1cm}$\qed
\end{pf*}

\begin{rem}
	The expressions derived in Theorem \ref{theorem33} are related to the Empirical Bayes hyperparameter estimator computations for regularized least-squares in \cite{chen2013implementation} and \cite{gonzalez2021noncausal}. In fact, in the absence of Lebesgue sampling, we would have $\bar{\mathbf{Q}}^{(j)}=\mathbf{z}_{1:N}\mathbf{z}_{1:N}^\top$, $\mathbf{C}^{(j)}=\mathbf{z}_{1:N}$, and the QR factorization in \eqref{qr} is now a thin QR factorization \cite[Thm 2.1.14]{horn2012} that provides alternative closed-form expressions for computing the hyperparameter estimator in one iteration using similar formulas to \eqref{lambdabetaopt} and \eqref{sigmaopt}. Contrary to the Riemann-sampling case, this work requires the EM algorithm to make the Empirical Bayes optimization tractable.
\end{rem}

\subsection{Transfer function description}
\label{subE}
The final theoretical contribution of this paper (Contribution C4) is the derivation of a more explicit expression for the estimated transfer function. Explicit expressions for general stable-spline kernels have been reported in \cite{scandella2022kernel} for unquantized output data with fully continuous-time inputs:
\begin{proposition}
	\label{prop2}
	The transfer function associated to the minimizer of \eqref{regularization2} can be written as
	\begin{equation}
		\label{transferfunction}
		\hat{G}(s) = \sum_{l=1}^{N} \hat{c}_l \hat{G}_l(s),
	\end{equation}
	where $\{\hat{c}_l\}_{l=0}^{N}$ is computed from \eqref{computec}, and 
	\begin{equation}
		\label{convolution}
		\hat{G}_l(s) = \int_{0}^{\infty} K(s;\tau) u(l\Delta-\tau) \mathrm{d}\tau, 
	\end{equation}
	with $K(s;\tau)$ being the Laplace transform of the kernel function $k(t,\tau)$.
\end{proposition}
\begin{pf*}{Proof.} 
	See \cite{scandella2022kernel}. \hfill $\hspace{1cm}$\qed
\end{pf*}
A similar expression to \eqref{transferfunction} also holds for this framework, as the only difference can be observed in the computation of the weights and the hyperparameters of the kernel (but not of the structure of the kernel itself). However, under the zero-order hold assumption on the input signal, we can provide an alternative representation of \eqref{transferfunction} for which the software implementation is easier and that does not rely on approximations of the intersample behavior of the input. This representation is stated in Lemma \ref{lemma35}.
\begin{lemma}
	\label{lemma35}
	Consider the optimization problem in \eqref{regularization2}, where $\mathcal{G}$ is the RKHS induced by a kernel $k$. The transfer function associated to the minimizer of \eqref{regularization2} can be written as
	\begin{equation}
		\hat{G}(s) = \hat{\mathbf{c}}^\top \bm{\Phi} \mathcal{K}(s), \notag
	\end{equation}
	where $\mathbf{\hat{c}}$ is computed from \eqref{computec}, $\bm{\Phi}$ is defined in \eqref{U}, and $\mathcal{K}(s)$ is a vector of size $N$ with entries $\mathcal{K}_{l}(s)$ given by the Laplace transform of the integrated kernel, i.e.,
	\begin{equation}
		\label{mathcalk}
		\mathcal{K}_{l}(s)=\int_{0}^\infty \left(\int_{\Delta [l-1]}^{\Delta l} k(t,\tau) \textnormal{d}\tau\right)e^{-st} \textnormal{d}t.
	\end{equation} 
\end{lemma}
\begin{pf*}{Proof.}
	See Appendix \ref{prooflemma35}. \hfill $\hspace{1cm}$\qed
\end{pf*}
\begin{corollary}
	\label{corollaryss}
	If the RKHS $\mathcal{G}$ is induced by the stable-spline kernel of order $q$, then the transfer function associated to the minimizer of \eqref{regularization2} can be written as $\hat{G}(s) = \hat{\mathbf{c}}^\top \bm{\Phi} \mathcal{K}(s)$, where $\mathbf{\hat{c}}$ is computed from \eqref{computec}, $\bm{\Phi}$ is defined in \eqref{U}, and $\mathcal{K}(s)$ is a vector of size $N$ with entries $\mathcal{K}_{l}(s)$ given by
	\begin{align}
		\mathcal{K}_{l}(s)&= \sum_{r=0}^{q-1} \frac{\gamma_{q,r}e^{-l\beta\Delta (2q-r-1)}(e^{\beta\Delta(2q-r-1)}-1)}{\beta(s+r\beta )(2q-r-1)} \notag \\
		&\hspace{-0.4cm}+\frac{(-1)^{q} \beta^{2q-1}e^{-l\Delta (s+\beta[2q-1])}(e^{\Delta(s+\beta[2q-1])}-1)}{(s+\beta[2q-1])\prod_{k=0}^{2q-1} (s+k\beta)}. \notag
	\end{align}
\end{corollary}
\begin{pf*}{Proof.}
	Direct from replacing $k(t,\tau)$ in \eqref{mathcalk} for \eqref{sq} and solving the integrals.  \hfill $\hspace{1cm}$\qed
\end{pf*}

In summary, the estimated transfer function of the Lebesgue-sampled continuous-time system of interest can be computed in a straightforward manner after the hyperparameter vector $\bm{\rho}$ and representer weighting vector $\hat{\mathbf{c}}$ are obtained. Both of these quantities have been proven to be computable from separate EM iterations in Theorems \ref{theorem33} and \ref{theorem31}, respectively.

\subsection{Algorithm}
To conclude this section, the full algorithm for non-parametric identification of Lebesgue-sampled continuous-time systems is described in Algorithm \ref{algorithm2}. For simplicity we replace the hyperparameter $\gamma$ for $\tilde{\gamma}$ in the description of the hyperparameter vector $\bm{\rho}$.

\begin{algorithm}
	\caption{Kernel-based non-parametric identification for Lebesgue-sampled continuous-time systems}
	\begin{algorithmic}[1]
		\State Input: $\mathbf{u}_{0:N-1},\mathcal{Y}_{1:N}$, initial hyperparameter estimate $\hat{\bm{\rho}}^{(1)}=[\hat{\tilde{\gamma}}^{(1)}, \hat{\bm{\beta}}^{(1)\top},{\hat{\sigma}^2{}}^{(1)}]^\top$, maximum number of MAP-EM iterations $M_{\textnormal{iter}}$
		\State Form $\bm{\Phi}$ as in \eqref{U}
		\For{$j = 1, 2, \dots,M_{\textnormal{iter}}$}
		\State Compute $\bar{\mathbf{Q}}^{(j)}$ from \eqref{secondmoment} using the minimax tilt-\hspace*{0.52cm}ing algorithm in \cite{botev2017normal}
		\State Factor $\bar{\mathbf{Q}}^{(j)}=\mathbf{C}^{(j)}{\mathbf{C}^{(j)}{}}^{\top}$ and $\mathcal{O}_{\hat{\bm{\beta}}^{(j)}}/\hat{\tilde{\gamma}}^{(j)} = \hspace*{0.43cm} \mathbf{L}_{\hat{\bm{\rho}}^{(j)}}\mathbf{L}_{\hat{\bm{\rho}}^{(j)}}^\top$
		\State Perform the QR factorization in \eqref{qr}
		\State Obtain $\hat{\bm{\rho}}^{(j+1)} = [\hat{\tilde{\gamma}}^{(j+1)}, \hat{\bm{\beta}}^{(j+1)},{\hat{\sigma}^2{}}^{(j+1)}]^\top$ \hspace*{0.43cm} from \eqref{lambdabetaopt} and \eqref{sigmaopt}
		\EndFor
		\State Compute initial estimate $\hat{\mathbf{c}}^{(1)}$ from the midpoint approximation in \eqref{approximatec}
		\For{$j = 1, 2, \dots,M_{\textnormal{iter}}$}
		\State Obtain $\hat{\mathbf{c}}^{(j+1)}$ from \eqref{ck1} with $\mathbf{K}$, $\tilde{\gamma}$ and $\tilde{\mathbf{z}}^{(j)}$ \hspace*{0.43cm} computing using $\hat{\bm{\rho}}^{(M_{\textnormal{iter}}}+1)$
		\EndFor
		\State Output: estimated transfer function $\hat{G}(s)={\hat{\mathbf{c}}^{(M_{\textnormal{iter}}+1)}{}}^{\top} \bm{\Phi} \mathcal{K}(s)$, with $\mathcal{K}(s)$ computed from \eqref{mathcalk} using $\hat{\bm{\beta}}^{(M_{\textnormal{iter}}+1)}$.		
	\end{algorithmic}
	\label{algorithm2}
\end{algorithm}
\begin{rem}
    Similarly as in lines 2 to 10 of Algorithm \ref{algorithm1}, instead of performing a fixed number of iterations, the iterations could be stopped after a stopping criterion is satisfied (line 6 in Algorithm~\ref{algorithm1}). In case of the loop in lines 3 to 8 in Algorithm~\ref{algorithm2}, this stopping criterion is defined as $\|\hat{\bm{\rho}}^{(j+1)}-\hat{\bm{\rho}}^{(j)}\|_2/\|\hat{\bm{\rho}}^{(j)}\|_2 <\epsilon$, while in case of the loop in lines 10 to 12 in Algorithm~\ref{algorithm2}, it is defined as $\|\hat{\mathbf{c}}^{(j+1)}-\hat{\mathbf{c}}^{(j)}\|_2/\|\hat{\mathbf{c}}^{(j)}\|_2 <\epsilon$, where the values of $\epsilon$ could be different in each stopping criteria.
\end{rem}

\section{Simulations}
\label{sec:simulations}
The performance of the novel non-parametric estimator is tested on a series of extensive Monte Carlo simulations. 

\subsection{Practically relevant example}
\label{sec:practicallyrelevant}
We consider a mass-spring-damper system with transfer function given~by
\begin{equation}
	\label{system1}
	G(s) = \frac{1}{\textnormal{m}s^2+\textnormal{d}s+\textnormal{k}},
\end{equation}
with mass $\textnormal{m}\hspace{-0.07cm}=\hspace{-0.07cm}0.05$[kg], damping coefficient $\textnormal{d}\hspace{-0.07cm}=\hspace{-0.07cm}0.2$[Ns/m], and spring constant $\textnormal{k}\hspace{-0.07cm}=\hspace{-0.07cm}1$[N/m]. The output is sensed with period $\Delta\hspace{-0.07cm}=\hspace{-0.07cm}0.1$[s], and $h\hspace{-0.07cm}=\hspace{-0.07cm}1$[m]. The input is a Gaussian white noise sequence of standard deviation $5[N]$ passed through a zero-order hold device with period $\Delta_u=3$[s]. The output prior to the Lebesgue sampling is computed using the \texttt{lsim} command in MATLAB with sampling time $0.1$[s], which delivers exact noiseless output values since the input is a zero-order hold signal. One hundred Monte Carlo runs are performed with a varying input and an additive Gaussian white noise prior to the Lebesgue sampling with standard deviation $0.05$[m]. Each run has a total time duration of $30$[s] (i.e., $300$ data points are sensed prior to Lebesgue sampling), and on average $N_{\textnormal{L}}=69$ output samples are obtained after sampling per run.

Three estimators are tested: the kernel-based continuous-time non-parametric estimator with equidistantly-sampled data \cite{pillonetto2010new,scandella2022kernel} using the stable-spline kernel of order 1 and the midpoint estimate $z(i\Delta)\approx \eta_i+h/2$ as output data ($\hat{g}_{\textnormal{rie}}$), this same estimator but using the noisy output $z(i\Delta)$ prior to Lebesgue sampling as output data ($\hat{g}_{\textnormal{or}}$), and the proposed approach (Algorithm \ref{algorithm2} of this paper, $\hat{g}_{\textnormal{leb}}$). Note that the oracle estimator $\hat{g}_{\textnormal{or}}$ cannot be implemented in practice, since we do not have direct knowledge of the system output before the event-sampler. This estimator is different from the commonly-denominated oracle estimator that uses the unattainable kernel $k(\tau,\xi)=g(\tau)g(\xi)$ \cite{chen2012estimation}. We measure the performance of each estimator with the fit metric
\begin{equation}
	\textnormal{fit} = 100 \left(1- \frac{\|\hat{\mathbf{x}}^j-\mathbf{x}\|_2}{\|\mathbf{x}-\bar{x}\mathbf{1}\|_2}\right), \notag
\end{equation}
where $\mathbf{x}$ is the noiseless output sequence (prior to Lebesgue sampling), $\hat{\mathbf{x}}^j$ is the simulated output sequence using the $j$th impulse response estimate, and $\bar{x}$ is the mean value of $\mathbf{x}$. The proposed estimator uses the stable-spline kernel of order 1 with a maximum number of EM iterations $M_{\textnormal{iter}}=40$, and $1000$ samples of a multivariate truncated Gaussian distribution are obtained to compute $\bar{\mathbf{Q}}^{\hspace{-0.02cm}(\hspace{-0.01cm}j\hspace{-0.01cm})}$\hspace{0.05cm}in\hspace{0.05cm}\eqref{secondmoment}.

\begin{figure}
	\centering{
		\includegraphics[width=0.47\textwidth]{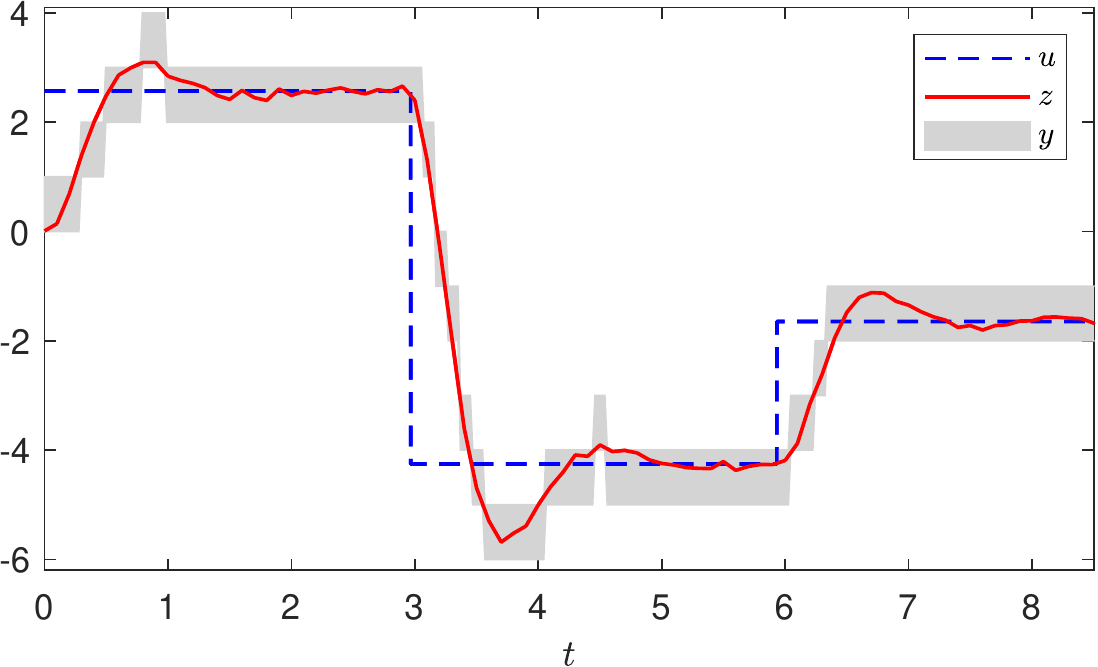}
		\vspace{-0.2cm}
		\caption{Input and output signals of the system \eqref{system1} corresponding to $8$[s] of one Monte Carlo run.}
		\label{fig2}}
\end{figure}

\begin{figure}
	\centering{
		\includegraphics[width=0.47\textwidth]{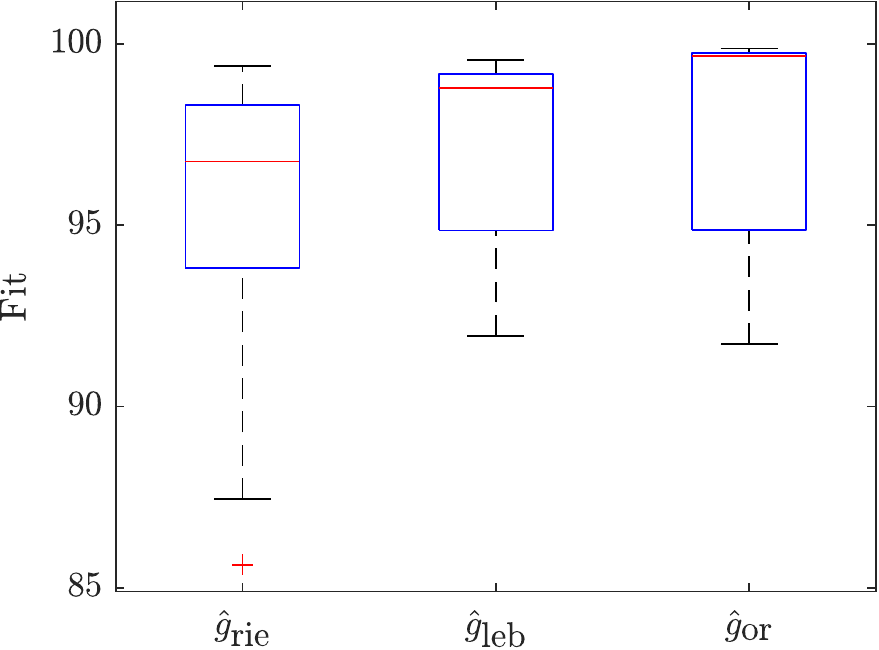}
		\vspace{-0.2cm}
		\caption{Boxplots of the fit metric for the case study,  Section \ref{sec:practicallyrelevant}. The Lebesgue-sampling-based estimator $\hat{g}_{\textnormal{leb}}$ achieves a better performance than the Riemann approach $\hat{g}_{\textnormal{rie}}$.}
		\label{fig3}}
\end{figure}

\begin{figure}
	\centering{
		\includegraphics[width=0.47\textwidth]{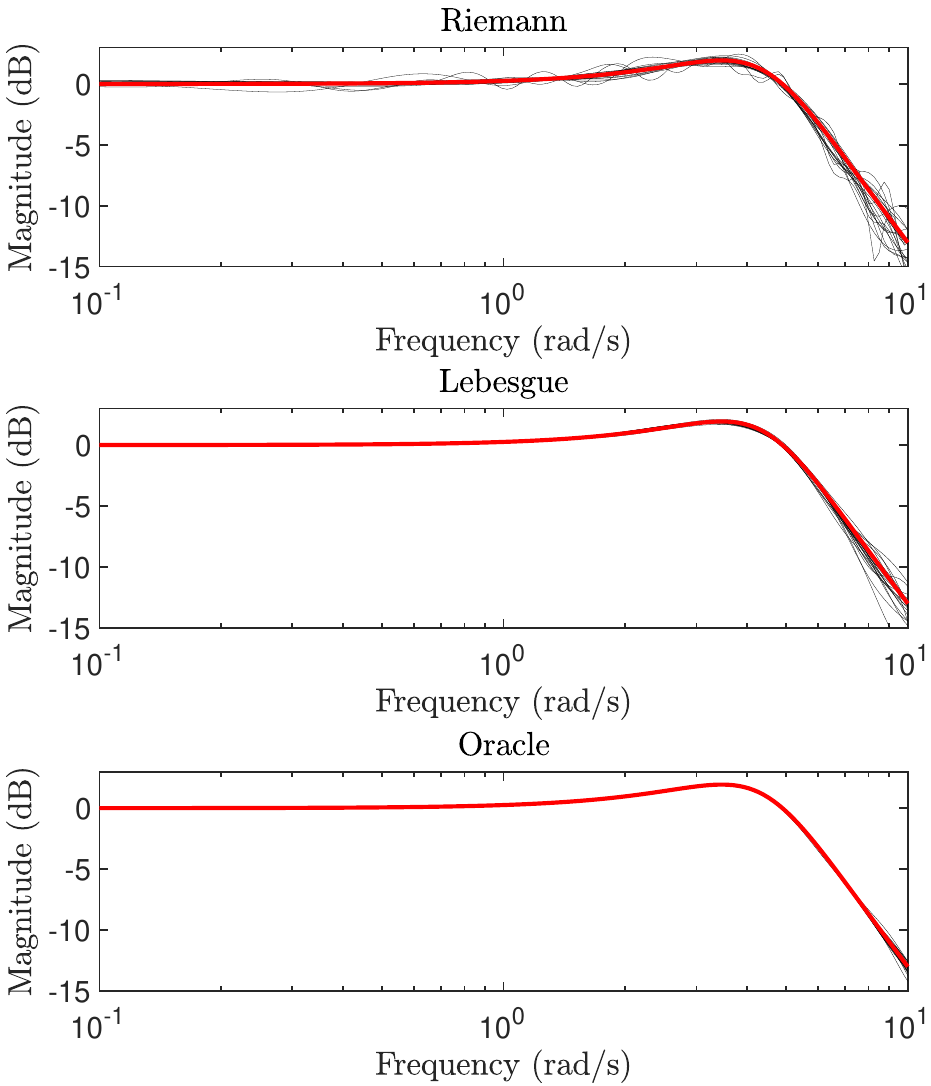}
		\vspace{-0.2cm}
		\caption{Bode magnitude plots of 20 Monte Carlo runs (black), compared to the true frequency response (red). Upper plot: equidistantly-sampled approach \cite{pillonetto2010new}; middle plot: proposed method; lower plot: oracle method (unattainable). The Bode plots of the Lebesgue-sampling approach, obtained via Lemma \ref{lemma35}, show much less variability than the Riemann approach over the Monte Carlo runs, and are comparable to the estimates produced by the oracle method.}
		\label{figbode}}
\end{figure}

A typical data set is shown in Figure \ref{fig2}. Note that the task of the proposed estimator is particularly challenging, since the overshoot of the output signal $z$ is rarely captured in the $y$ signal band due to the coarse grid produced by the threshold level $h$. To show the statistical performance of each estimator, we present the boxplots of the fit metric for each estimator in Figure \ref{fig3}. A graphical illustration of the proximity of the estimated frequency responses to the frequency response of the true system is presented in Figure \ref{figbode}, which shows 20 Bode magnitude plots of the frequency response estimates (obtained via Corollary \ref{corollaryss}) of each method, obtained from 20 noise realizations. As expected, the proposed approach achieves on average a better fit than the estimator that only uses the midpoint values $\eta_i+h/2$ as output. The $\hat{g}_{\textnormal{leb}}$ estimator is only slightly outperformed by the oracle estimator, despite having a low resolution for the output measurement mechanism and a $77\%$ reduction in output data samples on average.

\begin{rem}
An additional test has been conducted to assess the necessity of EM iterations for computing the optimal weights $\hat{\mathbf{c}}$. Under the same experimental conditions as above, we have compared the fit of the Lebesgue approach employing the initial estimate \eqref{approximatec} for the weight vector against the fit achieved with the estimator computed from the EM iterations outlined in Theorem \ref{theorem31}. We have observed that incorporating EM iterations for the weight vector has led to a better fit in 96 out of 100 Monte Carlo runs. This suggests that performing EM iterations for computing the weight vector is crucial for achieving the best performance.
    \end{rem}

\subsection{Effect of the threshold amplitude $h$}
\label{subsec:h}
The threshold amplitude plays an important role in the accuracy of any system identification method, since it is directly related to the size of the set uncertainty of the output measurement. The system in \eqref{system1} is identified under the same experimental conditions as Section \ref{sec:practicallyrelevant}, but now with $0.1[m]$ as standard deviation of the additive noise. Six different values of $h$ are tested, and for each value, one hundred Monte Carlo runs are recorded.

\begin{figure*}
	\centering{
		\includegraphics[width=0.87\textwidth]{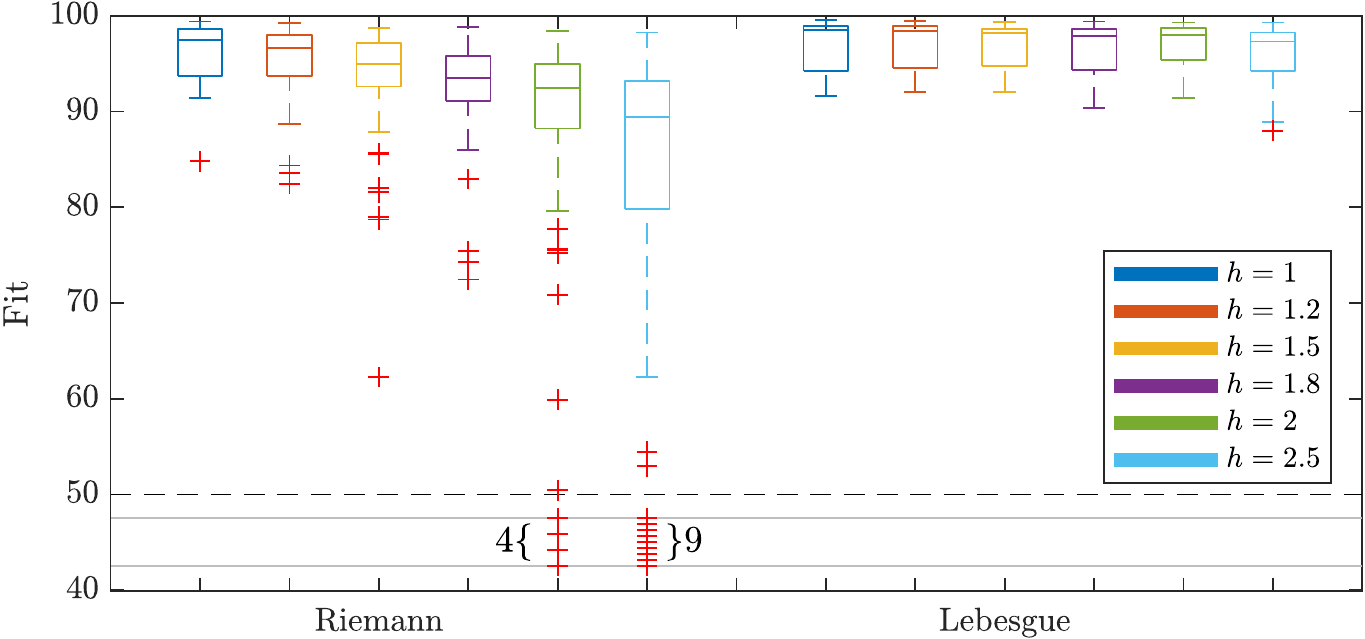}
		\vspace{-0.2cm}
		\caption{Boxplots of the fit metric for different values of threshold amplitude $h$, Section \ref{subsec:h}. Riemann sampling (left), Lebesgue sampling (right). While the estimator using the Riemann-sampling approach severely deteriorates its performance for coarser threshold grids, the proposed method produces excellent results for all values of $h$ in this study.}
		\label{fig4}}
\end{figure*}

The boxplots in Figure \ref{fig4} show that the performance of the standard (Riemann) non-parametric estimator severely deteriorates as the threshold amplitude $h$ grows. In sharp contrast, the proposed estimator remains accurate even when $h$ is large compared to the amplitude range of the unsampled output. In Table \ref{table1}, we have registered the average number of effective samples that are obtained for each simulation study. These numbers confirm the advantage of Lebesgue sampling over equidistant sampling in terms of resource efficiency, since the correct utilization of the set-uncertainty in the Lebesgue sampling strategy can lead to a sevenfold reduction in output data used in the identification process (from $300$ to $38.5$) with only minor performance detriment compared to Riemann sampling with $h=1$.

\begin{table}
	\caption{Average number of output samples retrieved from the Monte Carlo experiments for each threshold distance $h$. For reference, the number of samples for the equidistantly-sampled estimator is $300$.}
	\centering{\begin{tabular}{c|c|c|c|c|c|c}
				\label{table1}
			$h [m]$         & $1$    & $1.2$  & $1.5$  & $1.8$  & $2$    & $2.5$  \\ \hline
			Samples & $79.5$ & $69.2$ & $59.7$ & $51.7$ & $47.5$ & $38.5$
	\end{tabular}}
\end{table}

\subsection{Other benchmark systems}
\label{sec:benchmark}
\begin{figure*}
	\centering{
		\includegraphics[width=0.85\textwidth]{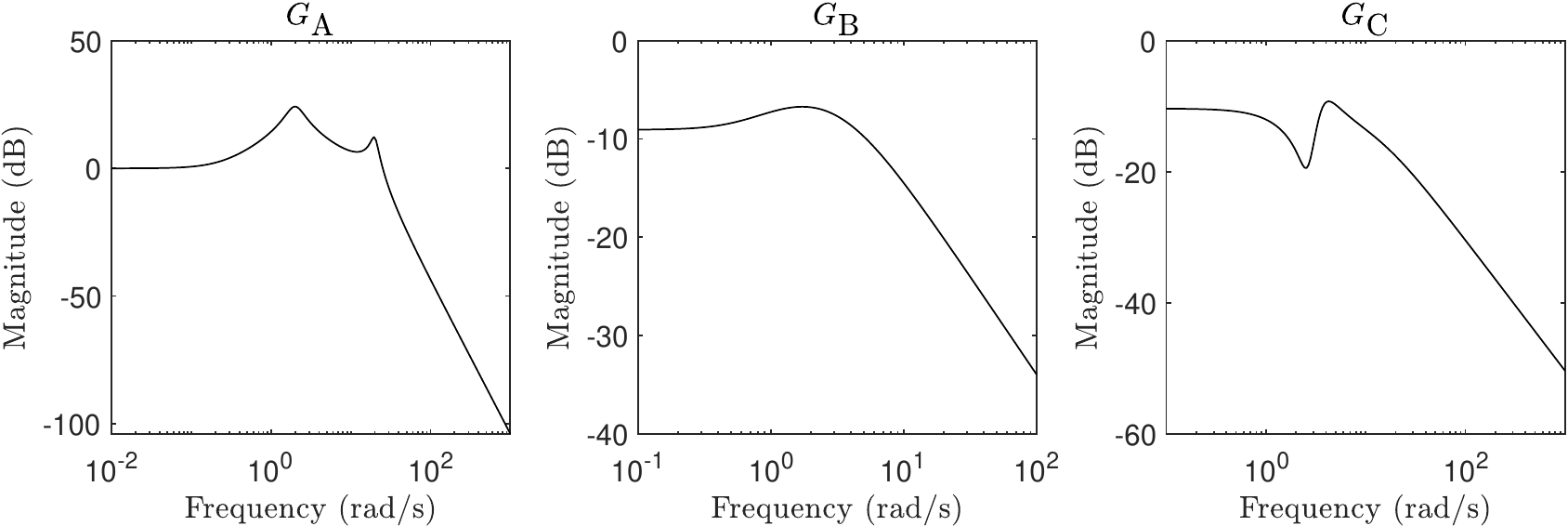}
		\vspace{-0.15cm}
		\caption{Bode magnitude plots of the three systems in Section \ref{sec:benchmark}. From left to right: $G_\textnormal{A}(s),G_\textnormal{B}(s)$ and $G_\textnormal{C}(s)$.}
		\label{fig5}}
\end{figure*}

\begin{figure*}
	\centering{
		\includegraphics[width=0.87\textwidth]{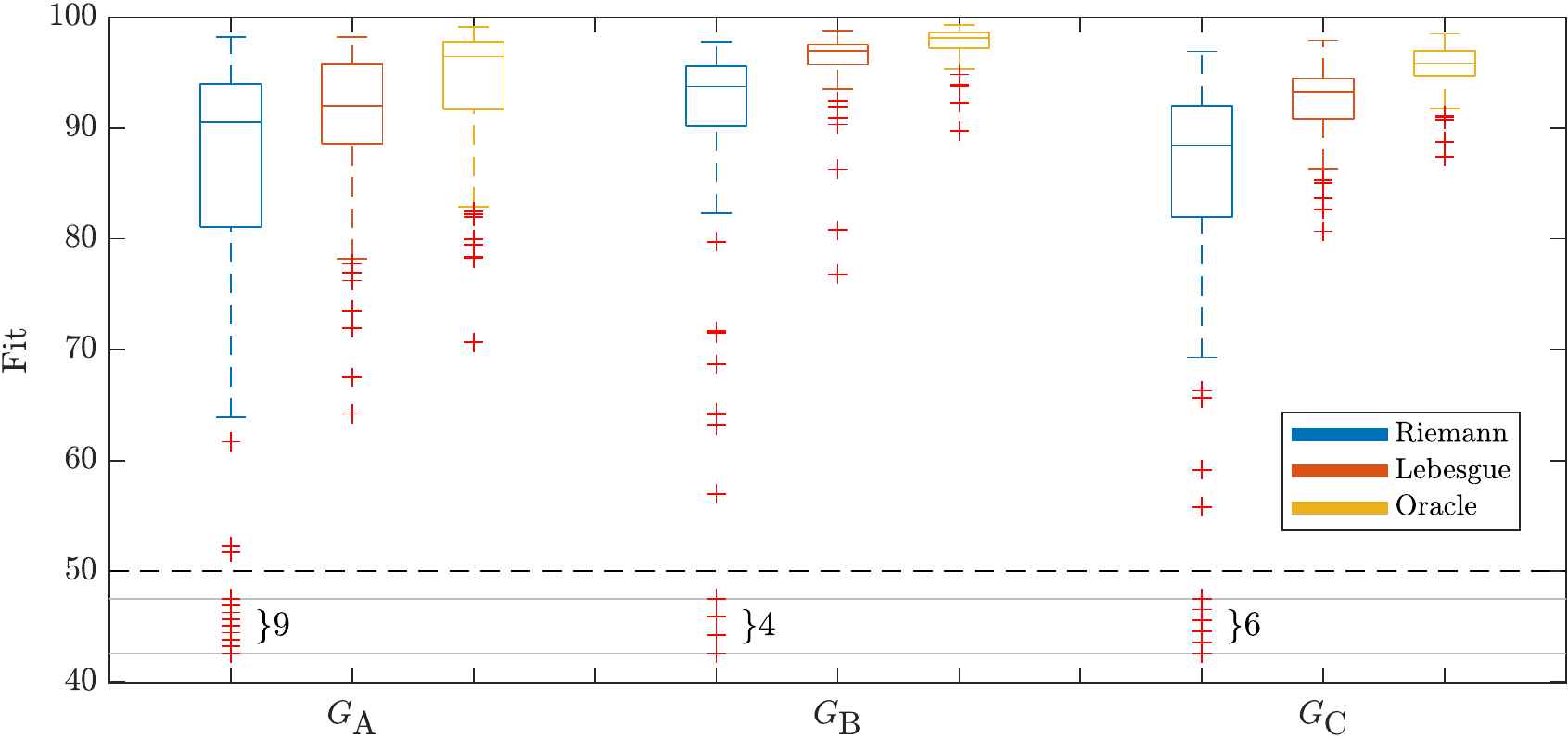}
		\vspace{-0.15cm}
		\caption{Boxplots of the fit metric for the three systems in Section \ref{sec:benchmark}. The proposed Lebesgue-sampling approach leads to an important gain in model fit compared to the Riemann approach in all the benchmark systems of this study.}
		\label{fig6}}
\end{figure*}

To show that the proposed estimator also performs well under different system setups, the next tests consider three more systems:
\begin{align}
	G_\textnormal{A}(s) \hspace{-0.02cm}&=\hspace{-0.02cm} \frac{-6400s+1600}{s^4+5s^3+408s^2+416s+1600}, \notag \\
	G_\textnormal{B}(s) \hspace{-0.02cm}&=\hspace{-0.02cm} \frac{27}{20}\frac{-2000s^3-3600s^2-2095s-396}{1350s^4\hspace{-0.03cm}+\hspace{-0.03cm}7695s^3\hspace{-0.03cm}+\hspace{-0.03cm}12852s^2\hspace{-0.03cm}+\hspace{-0.03cm}7796s\hspace{-0.03cm}+\hspace{-0.03cm}1520}, \notag \\
	G_\textnormal{C}(s) \hspace{-0.02cm}&=\hspace{-0.02cm} \frac{-3.025s^3\hspace{-0.03cm}-\hspace{-0.03cm}15.676s^2\hspace{-0.03cm}-\hspace{-0.03cm}32.802s\hspace{-0.03cm}-\hspace{-0.03cm}88.827}{s^4\hspace{-0.03cm}+\hspace{-0.03cm}16.52s^3\hspace{-0.03cm}+\hspace{-0.03cm}65.534s^2\hspace{-0.03cm}+\hspace{-0.03cm}235.01\hspace{-0.03cm}+\hspace{-0.03cm}292.948}, \notag
\end{align}
all of which have been used as benchmarks in other works on continuous-time system identification methods \cite{scandella2022kernel}. In particular, the Rao-Garnier system ($G_\textnormal{A}(s)$ in this work) has been tested in numerous works \cite{rao2002numerical,ljung2009experiments,garnier2015direct}, and is particularly challenging to identify due to its damped step response and stiffness. All systems have been excited by a Gaussian white noise of unit variance passed through a ZOH with period $\Delta_u=3[\textnormal{s}]$. The Bode plots of these systems are given in Figure \ref{fig5}, and the experimental conditions that are tested can be found in Table \ref{table2}, where we have also included the signal to noise ratio (SNR) between the output previous to Lebesgue sampling, $z$, and the additive noise, $v$. In Figure \ref{fig6}, we compare the fit metric of the proposed estimator to the Riemann and oracle estimators described in Section \ref{sec:practicallyrelevant} using 100 Monte Carlo runs. The results show that the Lebesgue sampling-based estimator outperforms the approach with equidistant sampling in all the systems considered in this study. Note that although the experimental conditions for $G_\textnormal{A}(s)$ give a better SNR, the performance is affected by a large threshold amplitude $h$ compared to the other cases. 

\begin{table}
	\caption{Experimental conditions for each system studied in Section \ref{sec:benchmark}.}
	\centering{\begin{tabular}{c|c|c|c|c}
	\label{table2}
			& $\Delta$    & $h$  & $\sigma$ & SNR [dB]\\ \hline
			$G_\textnormal{A}(s)$ & $0.01$ & $2.5$ & $0.3$ & $28.79$\\
			$G_\textnormal{B}(s)$ & $0.03$ & $0.2$ & $0.03$ & $22.82$\\
			$G_\textnormal{C}(s)$ & $0.03$ & $0.2$ & $0.03$ & $17.69$
	\end{tabular}}
\end{table}


\section{Conclusions}
\label{sec:conclusions}
The approach developed in this paper allows one to accurately identify Lebesgue-sampled systems based on input and output data. The main idea is to use all the available information for identification and control when dealing with Lebesgue-sampled signals. The proposed identification method, which is inspired by MAP estimation, kernel methods, and the EM algorithm, exploits the set uncertainty information in the output measurements to deliver more accurate models than the Riemann-sampling approach, while needing much fewer output samples. Thus, our method can enable systems with incremental encoders or with intermittent observations to be operated over less stringent sampling conditions (i.e., larger threshold amplitudes) without a severe loss in modeling accuracy. We have confirmed the advantages of the proposed algorithm in terms of statistical performance and resource efficiency in a series of extensive Monte Carlo simulations.

\section*{Acknowledgment}
This work is part of the research program VIDI with project number 15698, which is (partly) financed by the Netherlands Organization for Scientific Research (NWO).

\balance
\bibliographystyle{plain}        
\bibliography{References}     

\begin{thebibliography}{10}

\bibitem{aguero2017based}
J.~C. Ag{\"u}ero, K.~Gonz{\'a}lez, and R.~Carvajal.
\newblock {EM}-based identification of {ARX} systems having quantized output
  data.
\newblock {\em IFAC-PapersOnLine}, 50(1):8367--8372, 2017.

\bibitem{aravkin2014connection}
A.~Y. Aravkin, B.~M. Bell, J.~V. Burke, and G.~Pillonetto.
\newblock The connection between {B}ayesian estimation of a {G}aussian random
  field and {RKHS}.
\newblock {\em IEEE {T}ransactions on {N}eural {N}etworks and {L}earning
  {S}ystems}, 26(7):1518--1524, 2014.

\bibitem{arriagada2008relationship}
I.~A. Arriagada and J.~I. Yuz.
\newblock On the relationship between splines, sampling zeros and numerical
  integration in sampled-data models for linear systems.
\newblock In {\em 2008 American Control Conference}, pages 3665--3670. IEEE,
  2008.

\bibitem{aastrom2003systems}
K.~J. {\AA}str{\"o}m and B.~Bernhardsson.
\newblock Systems with {L}ebesgue sampling.
\newblock In {\em Directions in {M}athematical {S}ystems {T}heory and
  {O}ptimization}, pages 1--13. Springer, 2003.

\bibitem{bogachev1998gaussian}
V.~I. Bogachev.
\newblock {\em Gaussian Measures}.
\newblock Number~62. American Mathematical Society, 1998.

\bibitem{botev2017normal}
Z.~I. Botev.
\newblock The normal law under linear restrictions: simulation and estimation
  via minimax tilting.
\newblock {\em Journal of the Royal Statistical Society: Series B (Statistical
  Methodology)}, 79(1):125--148, 2017.

\bibitem{bottegal2017new}
G.~Bottegal, H.~Hjalmarsson, and G.~Pillonetto.
\newblock A new kernel-based approach to system identification with quantized
  output data.
\newblock {\em Automatica}, 85:145--152, 2017.

\bibitem{chen2013implementation}
T.~Chen and L.~Ljung.
\newblock Implementation of algorithms for tuning parameters in regularized
  least squares problems in system identification.
\newblock {\em Automatica}, 49(7):2213--2220, 2013.

\bibitem{chen2012estimation}
T.~Chen, H.~Ohlsson, and L.~Ljung.
\newblock On the estimation of transfer functions, regularizations and
  {G}aussian processes--{R}evisited.
\newblock {\em Automatica}, 48(8):1525--1535, 2012.

\bibitem{chen2012impulse}
T.~Chen, Y.~Zhao, and L.~Ljung.
\newblock Impulse response estimation with binary measurements: {A} regularized
  {FIR} model approach.
\newblock {\em IFAC Proceedings Volumes}, 45(16):113--118, 2012.

\bibitem{dempster1977maximum}
A.~P. Dempster, N.~M. Laird, and D.~B. Rubin.
\newblock Maximum likelihood from incomplete data via the {EM} algorithm.
\newblock {\em Journal of the Royal Statistical Society: Series B
  (Methodological)}, 39(1):1--22, 1977.

\bibitem{diao2018event}
J.-D. Diao, J.~Guo, and C.-Y. Sun.
\newblock Event-triggered identification of {FIR} systems with binary-valued
  output observations.
\newblock {\em Automatica}, 98:95--102, 2018.

\bibitem{dinuzzo2015kernels}
F.~Dinuzzo.
\newblock Kernels for linear time invariant system identification.
\newblock {\em SIAM Journal on Control and Optimization}, 53(5):3299--3317,
  2015.

\bibitem{dinuzzo2012representer}
F.~Dinuzzo and B.~Sch{\"o}lkopf.
\newblock The representer theorem for {H}ilbert spaces: a necessary and
  sufficient condition.
\newblock {\em Advances in Neural Information Processing Systems}, 25, 2012.

\bibitem{garnier2015direct}
H.~Garnier.
\newblock Direct continuous-time approaches to system identification.
  {O}verview and benefits for practical applications.
\newblock {\em European Journal of control}, 24:50--62, 2015.

\bibitem{garnier2014advantages}
H.~Garnier and P.~C. Young.
\newblock The advantages of directly identifying continuous-time transfer
  function models in practical applications.
\newblock {\em International Journal of Control}, 87(7):1319--1338, 2014.

\bibitem{godoy2011identification}
B.~I. Godoy, G.~C. Goodwin, J.~C. Ag{\"u}ero, D.~Marelli, and T.~Wigren.
\newblock On identification of {FIR} systems having quantized output data.
\newblock {\em Automatica}, 47(9):1905--1915, 2011.

\bibitem{gonzalez2021noncausal}
R.~A. Gonz{\'a}lez, C.~R. Rojas, and H.~Hjalmarsson.
\newblock Non-causal regularized least-squares for continuous-time system
  identification with band-limited input excitations.
\newblock In {\em Proceedings of the 60th IEEE Conference on Decision and
  Control}, pages 114--119, 2021.

\bibitem{gonzalez2021srivc}
R.~A. Gonz{\'a}lez, C.~R. Rojas, S.~Pan, and J.~S. Welsh.
\newblock The {SRIVC} algorithm for continuous-time system identification with
  arbitrary input excitation in open and closed loop.
\newblock In {\em Proceedings of the 60th IEEE Conference on Decision and
  Control}, pages 3004--3009, 2021.

\bibitem{gonzalez2023impulse}
R.~A. Gonz\'alez, K.~Tiels, and T.~Oomen.
\newblock Identifying {L}ebesgue-sampled continuous-time impulse response
  models: {A} kernel-based approach.
\newblock In {\em IFAC World Congress on Automatic Control,
  \textnormal{Yokohama, Japan}}, 2023.

\bibitem{horn2012}
R.~A. Horn and C.~R. Johnson.
\newblock {\em Matrix Analysis, \textnormal{2nd Edition}}.
\newblock Cambridge University Press, 2012.

\bibitem{kawaguchi2016system}
T.~Kawaguchi, S.~Hikono, I.~Maruta, and S.~Adachi.
\newblock System identification under {L}ebesgue sampling and its asymptotic
  property.
\newblock In {\em Proceedings of the 55th IEEE Conference on Decision and
  Control}, pages 2079--2084, 2016.

\bibitem{kimeldorf1970correspondence}
G.~S. Kimeldorf and G.~Wahba.
\newblock A correspondence between {B}ayesian estimation on stochastic
  processes and smoothing by splines.
\newblock {\em The Annals of Mathematical Statistics}, 41(2):495--502, 1970.

\bibitem{kon2021intermittent}
J.~Kon, N.~Strijbosch, S.~Koekebakker, and T.~Oomen.
\newblock Intermittent sampling in repetitive control: exploiting time-varying
  measurements.
\newblock In {\em Proceedings of the 60th IEEE Conference on Decision and
  Control}, pages 6566--6571, 2021.

\bibitem{liu2014survey}
Q.~Liu, Z.~Wang, X.~He, and D.~Zhou.
\newblock A survey of event-based strategies on control and estimation.
\newblock {\em Systems Science \& Control Engineering: An Open Access Journal},
  2(1):90--97, 2014.

\bibitem{ljung2009experiments}
L.~Ljung.
\newblock Experiments with identification of continuous time models.
\newblock In {\em 15th IFAC Symposium on System Identification,
  \textnormal{Saint Malo, France}}, volume~42, pages 1175--1180. Elsevier,
  2009.

\bibitem{mclachlan2007algorithm}
G.~J. McLachlan and T.~Krishnan.
\newblock {\em The {EM} {A}lgorithm and {E}xtensions}.
\newblock John Wiley \& Sons, 2007.

\bibitem{merry2013optimal}
R.~J.~E. Merry, M.~J.~G. van~de Molengraft, and M.~Steinbuch.
\newblock Optimal higher-order encoder time-stamping.
\newblock {\em Mechatronics}, 23(5):481--490, 2013.

\bibitem{piga2021learning}
D.~Piga, M.~Mejari, and M.~Forgione.
\newblock Learning dynamical systems from quantized observations: a {B}ayesian
  perspective.
\newblock {\em IEEE Transactions on Automatic Control}, 2021.

\bibitem{pillonetto2022regularized}
G.~Pillonetto, T.~Chen, A.~Chiuso, G.~De~Nicolao, and L.~Ljung.
\newblock {\em Regularized System Identification}.
\newblock Springer, 2022.

\bibitem{pillonetto2015tuning}
G.~Pillonetto and A.~Chiuso.
\newblock Tuning complexity in regularized kernel-based regression and linear
  system identification: {T}he robustness of the marginal likelihood estimator.
\newblock {\em Automatica}, 58:106--117, 2015.

\bibitem{pillonetto2010new}
G.~Pillonetto and G.~De~Nicolao.
\newblock A new kernel-based approach for linear system identification.
\newblock {\em Automatica}, 46(1):81--93, 2010.

\bibitem{pillonetto2014kernel}
G.~Pillonetto, F.~Dinuzzo, T.~Chen, G.~De~Nicolao, and L.~Ljung.
\newblock Kernel methods in system identification, machine learning and
  function estimation: A survey.
\newblock {\em Automatica}, 50(3):657--682, 2014.

\bibitem{pouliquen2016continuous}
M.~Pouliquen, A.~Goudjil, O.~Gehan, and E.~Pigeon.
\newblock Continuous-time system identification using binary measurements.
\newblock In {\em Proceedings of the 55th IEEE Conference on Decision and
  Control}, pages 3787--3792, 2016.

\bibitem{pouliquen2019identification}
M.~Pouliquen, E.~Pigeon, O.~Gehan, and A.~Goudjil.
\newblock Identification using binary measurements for {IIR} systems.
\newblock {\em IEEE Transactions on Automatic Control}, 65(2):786--793, 2019.

\bibitem{rao2002numerical}
G.~P. Rao and H.~Garnier.
\newblock Numerical illustrations of the relevance of direct continuous-time
  model identification.
\newblock In {\em 15th Triennial IFAC World Congress on Automatic Control,
  \textnormal{Barcelona, Spain}}, volume~35, pages 133--138, 2002.

\bibitem{risuleo2019identification}
R.~S. Risuleo, G.~Bottegal, and H.~Hjalmarsson.
\newblock Identification of linear models from quantized data: a
  midpoint-projection approach.
\newblock {\em IEEE Transactions on Automatic Control}, 65(7):2801--2813, 2019.

\bibitem{scandella2022kernel}
M.~Scandella, M.~Mazzoleni, S.~Formentin, and F.~Previdi.
\newblock Kernel- based identification of asymptotically stable continuous-time
  linear dynamical systems.
\newblock {\em International Journal of Control}, 95(6):1668--1681, 2022.

\bibitem{scholkopf2001generalized}
B.~Sch{\"o}lkopf, R.~Herbrich, and A.~J. Smola.
\newblock A generalized representer theorem.
\newblock In {\em International Conference on Computational Learning Theory},
  pages 416--426, 2001.

\bibitem{schoukens1994identification}
J.~Schoukens, R.~Pintelon, and H.~Van~Hamme.
\newblock Identification of linear dynamic systems using piecewise constant
  excitations: use, misuse and alternatives.
\newblock {\em Automatica}, 30(7):1153--1169, 1994.

\bibitem{strijbosch2019beyond}
N.~Strijbosch and T.~Oomen.
\newblock Beyond quantization in iterative learning control: Exploiting
  time-varying time-stamps.
\newblock In {\em IEEE American Control Conference (ACC)}, pages 2984--2989,
  2019.

\bibitem{strijbosch2022iterative}
N.~Strijbosch and T.~Oomen.
\newblock Iterative learning control for intermittently sampled data: Monotonic
  convergence, design, and applications.
\newblock {\em Automatica}, 139, \textnormal{Article 110171}, 2022.

\bibitem{wahba1990spline}
G.~Wahba.
\newblock {\em Spline Models for Observational Data}.
\newblock SIAM, 1990.

\bibitem{wu1983convergence}
C.~F.~J. Wu.
\newblock On the convergence properties of the {EM} algorithm.
\newblock {\em The {A}nnals of {S}tatistics}, pages 95--103, 1983.

\end{thebibliography}

\section*{Appendix}
\subsection{Proof of Lemma \ref{lemma32}}
\label{prooflemma32}
\begin{pf*}{Proof.}
	The MAP estimator for $\mathbf{c}$ is computed by
	\begin{align}
		\hat{\mathbf{c}}_{\textnormal{MAP}} &= \underset{\mathbf{c}\in \mathbb{R}^N}{\arg \max} \hspace{0.1cm}\textnormal{p}(\mathcal{Y}_{1:N}|\mathbf{c}) \hspace{0.03cm}\textnormal{p}(\mathbf{c}) \notag \\
		&=\hspace{-0.03cm}\underset{\mathbf{c}\in \mathbb{R}^N}{\arg \max} \bigg(\hspace{-0.03cm}-\frac{N}{2}\hspace{-0.02cm}\log(2\pi \sigma^2)\hspace{-0.03cm}-\hspace{-0.03cm}\frac{\log \det(2\pi \mathbf{K}^{-1}/\gamma)}{2}\notag \\
		\label{compareto}
		&\hspace{-0.07cm}+\hspace{-0.03cm} \sum_{i=1}^{N} \hspace{-0.02cm}\log\hspace{-0.04cm} \left[\hspace{-0.05cm}\int_{\eta_i}^{\eta_i\hspace{-0.03cm}+\hspace{-0.02cm}h}\hspace{-0.28cm} e^{\frac{-1}{2\sigma^2}\big(\hspace{-0.03cm}z_i\hspace{-0.03cm}-\hspace{-0.03cm}\mathbf{K}_i^\top \mathbf{c}\big)^{\hspace{-0.02cm}2}}\hspace{-0.06cm}\textnormal{d}z_i \hspace{-0.01cm}\right]\hspace{-0.08cm}-\hspace{-0.04cm}\frac{\gamma\mathbf{c}^{\hspace{-0.03cm}\top}\hspace{-0.02cm} \mathbf{Kc}}{2}\hspace{-0.03cm}\bigg),\hspace{-0.1cm}
	\end{align}
	where we have used the same derivation as for $\ell(\mathbf{x}_1)$ in Section \ref{section3A} for computing the log-likelihood term. By comparing \eqref{compareto} to \eqref{computec}, we find that $\hat{\mathbf{c}}$ in \eqref{computec} is simply the maximum\textit{ a posteriori} estimate of $\mathbf{c}$ within the model in \eqref{modelforc}. \hfill $\hspace{1cm}$\qed
\end{pf*}

\subsection{Proof of Theorem \ref{theorem31}}
\label{prooftheorem31}

\begin{pf*}{Proof.}
Since the $Q$ function provided by Lemma \ref{lemma33} is concave in $\mathbf{c}$, it is sufficient to obtain the point(s) which make the gradient of the objective function equal to zero. The gradient of $Q(\mathbf{c},\hat{\mathbf{c}}^{(j)})-\gamma\mathbf{c}^\top \mathbf{Kc}/2$ is given~by
\begin{align}
	&\frac{\partial}{\partial \mathbf{c}} \left( Q(\mathbf{c},\hat{\mathbf{c}}^{(j)})-\frac{\gamma\mathbf{c}^\top \mathbf{Kc}}{2}\right) \notag \\
	&= \frac{-1}{\sigma^2} \hspace{-0.03cm} \sum_{i=1}^{N} \hspace{-0.03cm} \int_{\eta_i}^{\eta_i+h}\hspace{-0.2cm} \mathbf{K}_i(\mathbf{K}_i^\top \hspace{-0.02cm} \mathbf{c}\hspace{-0.02cm}-\hspace{-0.02cm}z_i) \textnormal{p}(z_i|y(i\Delta),\hat{\mathbf{c}}^{(j)})\textnormal{d}z_i - \gamma\mathbf{Kc}\hspace{-0.02cm}. \notag
\end{align}
Setting the gradient to zero yields 
\begin{equation}
	\label{intermediate}
	\hat{\mathbf{c}}^{(j+1)}=\left(\sum_{i=1}^{N} \mathbf{K}_i \mathbf{K}_i^\top+ \gamma\sigma^2 \mathbf{K}\right)^{-1}\sum_{i=1}^{N} \mathbf{K}_i  \tilde{z}_i^{(j)},
\end{equation}
where we have defined the conditional mean $\tilde{z}_i^{(j)}$ as
\begin{equation}
	\label{intermediate0}
	\tilde{z}_i^{(j)} =  \int_{\eta_i}^{\eta_i+h} z_i \textnormal{p}(z_i|y(i\Delta),\hat{\mathbf{c}}^{(j)})\textnormal{d}z_i,
\end{equation}
and where we have used the fact that, for all $i=1,2,\dots, N$,
\begin{equation}
	\int_{\eta_i}^{\eta_i+h} \textnormal{p}(z_i|y(i\Delta),\hat{\mathbf{c}}^{(j)})\textnormal{d}z_i = 1. \notag
\end{equation}
The iterations in \eqref{ck1} are obtained from \eqref{intermediate} by rewriting the sum related to $\tilde{z}_i^{(j)}$ conveniently and using the fact that 
\begin{equation}
	\sum_{i=1}^{N} \mathbf{K}_i \mathbf{K}_i^\top = \mathbf{K}^2, \notag 
\end{equation}
which holds since $\mathbf{K}$ is symmetric. Finally, the explicit expression for $\tilde{z}_i^{(j)}$ in \eqref{tildezi} can be obtained directly from expanding the following alternative expression for \eqref{tildezi} based on applying Bayes' theorem on the conditional expectation in \eqref{intermediate0}:
\begin{equation}
	\tilde{z}_i^{(j)} = \frac{\int_{\eta_i}^{\eta_i+h}z_i \exp\left(\frac{-1}{2\sigma^2} [z_i-\mathbf{K}_i^\top \hat{\mathbf{c}}^{(j)} ]^2 \right) \textnormal{d}z_i}{\int_{\eta_i}^{\eta_i+h} \exp\left(\frac{-1}{2\sigma^2} [z_i-\mathbf{K}_i^\top \hat{\mathbf{c}}^{(j)} ]^2 \right) \textnormal{d}z_i}. \hspace{0.8cm}\qed  \notag
\end{equation}
\end{pf*}

\subsection{Proof of Theorem \ref{theorem32}}
\label{prooftheorem32}
\begin{pf*}{Proof.}
	We seek to derive the EM iterations for computing the maximum likelihood estimate in \eqref{eb}. By setting the latent variable as $\mathbf{z}_{1:N}$, we must compute the following $Q$ function
	\begin{equation}
		Q(\bm{\rho},\hat{\bm{\rho}}^{(j)}) = \mathbb{E}\left\{\log \textnormal{p}(\mathbf{z}_{1:N},\mathcal{Y}_{1:N}|\bm{\rho})|\mathcal{Y}_{1:N},\hat{\bm{\rho}}^{(j)}\right\}, \notag
	\end{equation}
	where it can be shown that (cf. Eq. (19) of \cite{godoy2011identification})
	\begin{equation}
		\textnormal{p}(\mathbf{z}_{1:N},\mathcal{Y}_{1:N}|\bm{\rho}) = \begin{cases}
			\textnormal{p}(\mathbf{z}_{1:N}|\bm{\rho}) &\textnormal{if } \mathbf{z}_{1:N} \in \mathcal{Y}_{1:N}, \\
			0 & \textnormal{otherwise,}
		\end{cases} \notag
	\end{equation}
	which, by exploiting \eqref{zrho}, leads to
	\begin{align}
		-2Q&(\bm{\rho},\hat{\bm{\rho}}^{(j)})= \notag \\
		&\log\det(2\pi\mathbf{S}_{\bm{\rho}})+\mathbb{E}\{\mathbf{z}_{1:N}^\top\mathbf{S}_{\bm{\rho}}^{-1}\mathbf{z}_{1:N} |\mathcal{Y}_{1:N},\hat{\bm{\rho}}^{(j)}\}. \notag
	\end{align}
	The iterations in \eqref{emiterationsthm32} follow from applying the commutativity property of the trace function to the expectation above.
	
	The minimization of $-2Q(\bm{\rho},\hat{\bm{\rho}}^{(j)})$ with respect to $\bm{\rho}$ provides the M-step of the EM iterations for computing a maximum of the likelihood of interest, which in turn is equivalent to solving (locally or globally) the optimization problem in \eqref{kernelopt}. \hfill $\hspace{1cm}$\qed
\end{pf*}

\subsection{Proof of Lemma \ref{lemmaO}}
\label{appendixlemmaO}
\begin{pf*}{Proof.}
	Consider the zero-order hold representation (valid for $t\in[0,\Delta N)$),
	\begin{equation}
		\label{uzoh}
		u(t) = \sum_{k=0}^{N-1} u(k\Delta) \mathbbm{1}(\Delta k\leq t < \Delta [k+1]),
	\end{equation}
	with $\mathbbm{1}(\cdot)$ being the indicator function (i.e., $1$ if $(\cdot)$ is satisfied, and $0$ otherwise). Thus, we compute
	\begin{align}
		&u(i\Delta-\xi) u(j\Delta-\tau) = \sum_{k=0}^{N-1} \sum_{l=0}^{N-1} u(k\Delta) u(l\Delta) \notag \\
		\label{us}
		&\hspace{-0.07cm}\times\hspace{-0.04cm} \mathbbm{1}\hspace{-0.02cm}(\Delta [i\hspace{-0.05cm}-\hspace{-0.05cm}k\hspace{-0.05cm}-\hspace{-0.05cm}1]\hspace{-0.05cm}< \hspace{-0.05cm}\tau\hspace{-0.05cm} \leq\hspace{-0.06cm} \Delta[i\hspace{-0.05cm}-\hspace{-0.05cm}k] \hspace{-0.04cm}\wedge \hspace{-0.04cm}\Delta [j\hspace{-0.05cm}-\hspace{-0.05cm}l\hspace{-0.06cm}-\hspace{-0.06cm}1]\hspace{-0.04cm}< \hspace{-0.05cm}\xi\hspace{-0.05cm} \leq \hspace{-0.05cm} \Delta [j\hspace{-0.05cm}-\hspace{-0.05cm}l]).
	\end{align}
	Since the integral of interest ranges from $0<\tau,\xi<\infty$, the elements of \eqref{us} for $i-k\leq 0$ and $j-l\leq 0$ can be discarded. In other words, within the domain of integration, we can write $u(i\Delta-\xi) u(j\Delta-\tau)$ as \eqref{us} but with summation upper limits $i-1$ and $j-1$ instead of $N-1$, respectively. Thus, interchanging summation and integration yields
	\begin{equation}
		\mathbf{K}_{ij} \hspace{-0.04cm}=\hspace{-0.04cm} \sum_{k=0}^{i-1} \hspace{-0.02cm}\sum_{l=0}^{j-1} \hspace{-0.04cm}u(\hspace{-0.01cm}k\Delta\hspace{-0.01cm}) u(\hspace{-0.01cm}l\Delta\hspace{-0.01cm}) \hspace{-0.04cm}\int_{\hspace{-0.03cm}\Delta [i\hspace{-0.01cm}-\hspace{-0.01cm}k\hspace{-0.01cm}-\hspace{-0.01cm}1]}^{\Delta[i-k]}\hspace{-0.02cm}\int_{\hspace{-0.03cm}\Delta [j\hspace{-0.01cm}-\hspace{-0.01cm}l\hspace{-0.01cm}-\hspace{-0.01cm}1]}^{\Delta[j-l]} \hspace{-0.05cm}k(\xi,\hspace{-0.02cm}\tau) \textnormal{d}\tau \textnormal{d}\xi. \notag
	\end{equation}	
	Alternatively, we can write this entry of the kernel matrix as $\mathbf{U}_j^\top \mathcal{O}_{\bm{\beta}} \mathbf{U}_i$, where $\mathbf{U}_j$ and $\mathbf{U}_i$ are the $j$th and $i$th columns of $\bm{\Phi}^\top$, respectively, and $\mathcal{O}_{\bm{\beta}}$ has entries that are given by \eqref{Oij}. Since $\mathcal{O}_{\bm{\beta}}$ does not depend on $i$ nor $j$, it is possible to describe the complete matrix $\mathbf{K}$ by stacking the column vectors $\mathbf{U}_j$ and $\mathbf{U}_i$, leading to \eqref{Kdecomposition}. \hfill $\hspace{1cm}$\qed
\end{pf*}	

\subsection{Proof of Theorem \ref{theorem33}}
\label{prooftheorem33}

\begin{pf*}{Proof.}
	Let us first rewrite the $\log \det$ term in \eqref{emiterationsthm32}. Thanks to the Weinstein–Aronszajn identity \cite[1.3.P28]{horn2012}, we have
	\begin{align}
		\log\det(\mathbf{S}_{\bm{\rho}}) &= \log \det (\sigma^2 \mathbf{I})+\log \det (\bm{\Phi} \mathbf{L}_{\bm{\rho}}\mathbf{L}_{\bm{\rho}}^\top \bm{\Phi}^\top + \mathbf{I}) \notag \\
		&= N \log \sigma^2 + \log \det(\mathbf{L}_{\bm{\rho}}^\top \bm{\Phi}^\top \bm{\Phi} \mathbf{L}_{\bm{\rho}}+ \mathbf{I}) \notag \\
		\label{logdetS}
		&= N \log \sigma^2 + 2\log\det(\mathbf{R}_{1,\bm{\rho}}),
	\end{align}
	where the identity in \eqref{R1R1} has been used in the last step. We now study the trace term in \eqref{emiterationsthm32}. Note that, by the matrix inversion lemma,
	\begin{equation}
		\mathbf{S}_{\bm{\rho}}^{-1} = \sigma^{-2}\mathbf{I}-\sigma^{-2} \bm{\Phi} \mathbf{L}_{\bm{\rho}} (\mathbf{L}_{\bm{\rho}}^\top \bm{\Phi}^\top \bm{\Phi} \mathbf{L}_{\bm{\rho}} + \mathbf{I})^{-1} \mathbf{L}_{\bm{\rho}}^\top \bm{\Phi}^\top, \notag
	\end{equation}
	which leads to
	\begin{align}
		\textnormal{tr}\{\mathbf{S}_{\bm{\rho}}^{-1}\bar{\mathbf{Q}}^{(j)}\}&=\textnormal{tr}\{{\mathbf{C}^{(j)}{}}^{\top}\mathbf{S}_{\bm{\rho}}^{-1} \mathbf{C}^{(j)}\} \notag \\
		&\hspace{-1.89cm}= \hspace{-0.08cm}\frac{\textnormal{tr}\{\hspace{-0.02cm}\bar{\mathbf{Q}}^{\hspace{-0.02cm}(\hspace{-0.01cm}j\hspace{-0.01cm})}\hspace{-0.04cm}\}}{\sigma^2} \hspace{-0.04cm} - \hspace{-0.04cm} \frac{\textnormal{tr}\{\hspace{-0.02cm}{\mathbf{C}^{\hspace{-0.02cm}(\hspace{-0.01cm}j\hspace{-0.01cm})}{}}^{\hspace{-0.07cm}\top}\hspace{-0.03cm}\bm{\Phi} \mathbf{L}_{\hspace{-0.02cm}\bm{\rho}}\hspace{-0.02cm} (\hspace{-0.02cm}\mathbf{L}_{\hspace{-0.02cm}\bm{\rho}}^{\hspace{-0.02cm}\top} \bm{\Phi}^{\hspace{-0.04cm}\top} \hspace{-0.05cm} \bm{\Phi} \mathbf{L}_{\bm{\rho}} \hspace{-0.04cm}+\hspace{-0.04cm} \mathbf{I})^{\hspace{-0.02cm}-\hspace{-0.02cm}1} \hspace{-0.03cm}\mathbf{L}_{\bm{\rho}}^\top \hspace{-0.04cm}\bm{\Phi}^{\hspace{-0.04cm}\top} \mathbf{C}^{\hspace{-0.02cm}(\hspace{-0.01cm}j\hspace{-0.01cm})}\hspace{-0.02cm}\}}{\sigma^2}. \notag
	\end{align}
	This expression, when written in terms of $\mathbf{R}_{1,\bm{\rho}}$ and $\mathbf{R}_{2,\bm{\rho}}$ via \eqref{R1R1} and \eqref{R1R2}, is simply 
	\begin{align}
		\textnormal{tr}\{\mathbf{S}_{\bm{\rho}}^{-1}\bar{\mathbf{Q}}^{(j)}\}&= \frac{\textnormal{tr}\{\bar{\mathbf{Q}}^{(j)}-\mathbf{R}_{2,\bm{\rho}}^\top \mathbf{R}_{2,\bm{\rho}}\}}{\sigma^2} \notag \\
		\label{trSQ}
		&= \frac{\|\mathbf{C}^{(j)}\|_{\textnormal{F}}^2 -\|\mathbf{R}_{2,\bm{\rho}}\|_{\textnormal{F}}^2}{\sigma^2},
	\end{align}
	where we have used the definition of the Frobenius norm in the last line. By combining the results in \eqref{logdetS} and \eqref{trSQ}, we reach
	\begin{align}
		\hat{\bm{\rho}}^{(j+1)} &= \underset{\bm{\rho} \in \bm{\Gamma}}{\arg \min} \bigg( N \log \sigma^2 \notag \\
		\label{concentrate}
		&+ 2\log \det(\mathbf{R}_{1,\bm{\rho}})+\frac{\|\mathbf{C}^{(j)}\|_{\textnormal{F}}^2 -\|\mathbf{R}_{2,\bm{\rho}}\|_{\textnormal{F}}^2}{\sigma^2}\bigg).
	\end{align}
	Since both $\mathbf{R}_{1,\bm{\rho}}$ and $\mathbf{R}_{2,\bm{\rho}}$ depend on $\mathbf{L}_{\bm{\rho}}$, which in turn is already factored by a scalar variable $1/\gamma$, the dependence on $\sigma$ in the $\mathbf{R}$ matrices is redundant for the optimization above. Therefore, we can concentrate the cost function by minimizing \eqref{concentrate} over $\sigma^2$ first, which leads to \eqref{sigmaopt}. Replacing \eqref{sigmaopt} in \eqref{concentrate} and neglecting constant terms leads to \eqref{lambdabetaopt}, which concludes the proof. \hfill $\hspace{1cm}$\qed
\end{pf*}

\subsection{Proof of Lemma \ref{lemma35}}
\label{prooflemma35}
\begin{pf*}{Proof.}
	Under the zero-order hold intersample behavior assumption, the input description in \eqref{uzoh} permits rewriting the convolution in \eqref{convolution} as
	\begin{equation}
		\hat{G}_l(s) = \sum_{k=0}^{l-1} u(k\Delta) \int_{\Delta[l-k-1]}^{\Delta[l-k]} K(s;\tau)\textnormal{d}\tau. \notag
	\end{equation}
	Interchanging the integrals above leads to $\hat{G}_l(s)$ being equal to the $l$th row of $\bm{\Phi}$ multiplied by $\mathcal{K}$ defined in \eqref{mathcalk}. This fact, together with the representer theorem description \eqref{transferfunction}, leads to the desired result. \hfill $\hspace{1cm}$\qed
\end{pf*}      

\end{document}